\documentclass[useAMS,usenatbib,fleqn]{mnras}
\usepackage{graphicx,amssymb, txfonts}
\usepackage[dvips]{epsfig}
\usepackage{amssymb}
\usepackage{lscape}
\usepackage{float}
\usepackage{color}
\usepackage{graphicx}
\usepackage{verbatim}
\usepackage{rotating}
\usepackage{ulem}

\def\pb{Pa$\beta$}
\def\brgamma{\rm Br$\gamma$}

\def\feii{[Fe\,{\sc ii}]}
\def\feiiumdois{[Fe\,{\sc ii}]\,$\lambda\,1.2570\,\mu$m}

\def\pii{[P\,{\sc ii}]}
\def\piium{[P\,{\sc ii}]\,$\lambda\,1.1886\,\mu$m}

\def\sii{[S\,{\sc ii}]}

\def\hii{H\,{\sc ii}}
\def\nii{[N\,{\sc ii}]}

\def\h2{H$_2$}
\def\H212{H$_2$\,$\lambda\,2.1218\,\mu$m}
\def\p1{Paper~I}

\def\kms {$\rm km\,s^{-1}$}

\def\msun{\,M$_{\odot}$}
\def\yr{\,yr$^{-1}$}

\def\xy{x$_{\rm y}$}
\def\xyi{x$_{\rm yi}$}
\def\xio{x$_{\rm io}$}
\def\xo{x$_{\rm o}$}

\def\my{m$_{\rm y}$}
\def\myi{m$_{\rm yi}$}
\def\mio{m$_{\rm io}$}
\def\mo{m$_{\rm o}$}
\def\starlight{{\sc starlight}}
\def\profit{{\sc profit}}

\title[Outflows, Inflows and Star Formation in NGC\,2110]{Outflows, Inflows and Young Stars in the inner 200~pc of the Seyfert galaxy  NGC\,2110 }

\author[Diniz et al.]{Marlon R. Diniz$^{1}$\thanks{E-mail:
diniz.mr@gmail.com}, Rogemar A. Riffel$^{1,2}$,  Thaisa Storchi-Bergmann$^{3}$,  Rog\'erio Riffel$^{3}$%
\\
$^{1}$ Universidade Federal de Santa Maria, Departamento de F\'isica, Centro de Ci\^encias Naturais e Exatas, Santa Maria, RS, 97105-900, Brazil\\
$^2$ Johns Hopkins University, Department of Physics \& Astronomy, Bloomberg Center, 3400 N. Charles St, Baltimore, MD 21218, USA\\
$^{3}$ Universidade Federal do Rio Grande do Sul, Instituto de F\'isica, CP 15051, Porto Alegre,  RS, 91501-970, Brazil
}

\begin{document}

%\date{Accepted 1988 December 15. Received 1988 December 14; in original form 1988 October 11}

\pagerange{\pageref{firstpage}--\pageref{lastpage}} \pubyear{2019}

\maketitle

\label{firstpage}
\begin{abstract}
We present a two-dimensional mapping of stellar population age components, emission-line fluxes, gas excitation and kinematics within the inner $\sim200$\,pc  of the Seyfert 2 galaxy NGC\,2110. We used the Gemini North Integral Field Spectrograph (NIFS) in the J and K bands at a spatial resolution of $\sim22$\,pc. The unresolved nuclear continuum is originated in combined contributions of young stellar population (SP; age$\leq100$\,Myr), a featureless AGN continuum and hot dust emission. The young-intermediate SP  ($100<$age$\leq700$~Myr) is distributed in a ring-shaped structure at $\approx140$\,pc from the nucleus, which is roughly coincident with the lowest values of the stellar velocity dispersion. In the inner $\approx115$\,pc the old SP (age$>2$~Gyr) is dominant.
The \feiiumdois~emission-line flux distribution is correlated with the radio emission and its kinematics comprise two components, one from gas rotating in the galaxy plane and another from gas in outflow within a bicone oriented along north--south. These outflows seem to originate in the interaction of the radio jet with the ambient gas producing shocks that are the main excitation mechanism of the \feii\ emission. We estimate: (1) an ionized gas mass outflow rate of $\sim0.5$~\msun\yr\ at $\sim$70~pc from the nucleus; and (2) a kinetic power for the outflow of only 0.05\% of the AGN bolometric luminosity implying weak feedback effect on the galaxy.\\
\hfill{\bf Keywords}: galaxies: individual (NGC\,2110) -- galaxies: Seyfert -- infrared: galaxies -- galaxies: stellar populations -- galaxies: kinematics and dynamics
\end{abstract}

\section{Introduction}

Detailed studies on the gas kinematics and excitation in the inner region ($<1$\,kpc) of active galaxies are necessary to better understand feeding and feedback processes of Active Galactic Nuclei (AGNs) and their effect on galaxy evolution. In addition, the mapping of the stellar populations age distributions in the vicinity of AGN is fundamental to investigate a possible relation between AGN activity and star formation. 

In recent years, theoretical and observational studies have been carried out to investigate how the gas is transferred from galactic to nuclear scales \citep[e.g.][]{knapen2000,emsellem01,mw02,crenshaw2003,fathi2005,rogemar2008,davies09,schartmann2010,hopkins2012,schartmann2014}. For a recent review on observational constraints on the feeding of supermassive black holes, see \citet{thaisa19}. 
Gas inflows can either trigger star formation in the circumnuclear region of the galaxy and/or initiate an episode of nuclear activity. In fact, circumnuclear star forming regions are  often observed in active galaxies \citep[e.g.][]{morgan58,sersic67,imanishi2000,ardila2003,rogerio07,rogemar2009b,rogemar16,mito2018}.

For luminous AGN ($L_{\rm bol}\gtrsim10^{43}$ erg\,s$^{-1}$), the feedback processes -- e.g. effect of the radiation and winds emitted from accretion disk on the host galaxy --  are  claimed to play an important role in the evolution of galaxies \citep{bower2006,somerville2008}.  However, for lower luminosity objects the effect of the AGN in the host galaxy is  still an open question. Comparing the gas kinematics to the stellar kinematics, on scales from tens to hundreds parsecs, allows the mapping of AGN winds and their effect on the host galaxy properties.

One way to verify how this interaction occurs is through the study of the excitation and kinematics of the gas in the Narrow Line Region (NLR). Using the emission lines in the near-infrared (NIR), such as \feii$\lambda\lambda 1.257, 1.644~\mu$m, \pb, \brgamma~and \H212, one can map the main mechanisms of gas excitation as well as the kinematics of different gas phases. 
While some works point out a correlation between the \feii~and the radio emission, indicating that the \feii~emission in AGNs is originated by shocks due to radio jets \citep{forbes1993,blietz1994}, others argue that its emission is due to photoionization by the nuclear source \citep{simpson96,das2005,das2006}.

The mapping and characterization of the stellar populations (SPs) in active galaxies is needed in order to investigate the relation between nuclear activity and star formation, providing ingredients -- such as star formation history (SFH) and star formation rate (SFR) -- to constrain galaxy evolution models.

An advantage of using the NIR instead of optical observations is that the former allows the study of more obscured regions. This is due to the fact that the extinction is smaller in the NIR than in optical bands, although the detection of spectral features of young stars in this spectral range is not an easy task \citep{maraston05,rogerio07,rogerio2008,martins13,zibetti2013,rogerio2015,mito2018}.

By performing stellar population synthesis using models of evolutionary population synthesis (EPS), which include a suitable treatment of stars in the Thermally Pulsing Asymptotic Giant Branch (TP-AGB) phase \citep{maraston05}, it is possible to study in detail the properties of the stellar populations that dominate the NIR emission. This is due to the fact that these models include the contributions of molecules such as CN and TiO, characteristic of stars in this evolutionary phase. 

Our group AGNIFS (for AGN Integral Field Spectroscopy), uses NIR data obtained with the Gemini NIFS instrument at the Gemini North Telescope. We have been finding, as a frequent characteristic of the circumnuclear SP of AGN host galaxies, an important contribution of an intermediate age ($\sim700$\,Myr) component  \citep{rogemar2010b,rogerio2011c,ngc1068,astor17,diniz17,luis19,natacha19}.

In this work, we use NIFS observations to map the gas excitation and kinematics, as well as the stellar populations, of the inner 200~pc radius of NGC\,2110, an SO galaxy hosting a Sy\,2 nucleus and located at a distance of 30.2 Mpc (from NED\footnote{NASA/IPAC extragalactic database}, derived assuming a redshift of 0.007789 and $H_0=73$ km\,s$^{-1}$\,Mpc$^{-1}$), for which 1$^{\prime\prime}$ corresponds to 146\,pc.  In a previous study \citep{diniz15}, we used observations in the K band to map the molecular and ionized gas kinematics.
%. By comparing the molecular gas and stellar velocity fields, 
The flux distributions of \h2~and \brgamma~emission lines were found to be distinct: while the \h2~emission is observed over the whole field of view ($3^{\prime\prime}\times3^{\prime\prime}$), the \brgamma~emission is collimated along PA$\approx-30^{\circ}$.
 The gas kinematics revealed inflows towards the nucleus along spiral arms as well as a compact outflow within the inner $\approx70$~pc. The derived mass inflow rate and outflow rates in warm molecular gas (that originates the K-band H$_2$ emission) are $\approx5\times10^{-4}$~\msun\yr~and $\approx4\times10^{-4}$~\msun\yr, respectively. 
%The \h2~velocity dispersion results show that the lowest values seems to be forming a ring-like structure around the nucleus. Suggesting that a young/intermediate stellar population might be co-spatially with these low $\sigma$ values. 
We concluded that the main \h2~excitation mechanism is heating of the gas by X-rays emitted from the AGN. 

\citet{durre14} have also recently presented NIR IFS of the central region of NGC\,2110 using J-band data obtained with the Keck I OSIRIS instrument and reported the detection of star clusters along with a 90$\times$35~pc bar of shocked gas around the nucleus of NGC\,2110.  They derived a star formation rate of 0.3\,\msun\yr~and argue that the star formation is triggered by the AGN feedback.

Optical IFS of the inner $1.1\times1.6$~kpc$^2$ ($7''\times10''$) obtained with the Gemini GMOS instrument by \citet{allan2014}, revealed 
%that the ionized gas kinematics is complex, presenting  
four kinematic components:  (i) a cold gas disk (with velocity dispersion ($\sigma$) in the range 60--90~\kms) consistent with the H$_2$ kinematics of \citet{diniz15},  (ii) a warm gas disk ($\sigma=100-220$~\kms), (iii) a nuclear outflow ($\sigma=220-600$~\kms) and (iv) a cloud  ($\sigma=60-80$~\kms), which was identified as ``the northen cloud''.  
In agreement with \citet{diniz15}, these authors identified both inflowing and outflowing gas. They derived a ionized gas mass inflow and outflow rates of $\approx2\times10^{-2}$~\msun\yr~and $\approx0.9$~\msun\yr, respectively.

This paper is organized as follows: in Section 2 we describe the observations and data reduction procedures,  Section 3 presents two-dimensional maps for the emission-line fluxes, gas kinematics, flux ratios and contributions of  the stellar populations to the observed continuum emission. The results are discussed in  Section 4 and Section 5 summarizes the conclusions of this work.

\section{Observations, data reduction and analysis}\label{obs}

Two-dimensional NIR spectroscopic data were obtained at the Gemini North Telescope with the Near Infrared Integral Field Spectrograph \citep[NIFS; ][]{mcgregor03}, under the programmes GN-2010B-Q-25 and GN-2015B-Q-29. NGC\,2110 is part of the sample of the Brazilian Gemini Large Program  ``NIFS survey of feeding and feedback processes in nearby active galaxies" (P.I. Storchi-Bergmann) described in \citet{rogemar2018} and aimed to obtain J and K bands IFS of a sample of 20 nearby AGNs.  NIFS has a square field of view of  3$^{\prime\prime}\times$3$^{\prime\prime}$, which corresponds to $\sim440\times440$\,pc$^{2}$ at NGC\,2110.

The data reduction was accomplished using tasks contained in the {\sc nifs} package, which are part of {\sc gemini iraf} package, as well as standard {\sc iraf} tasks and Interactive Data Language (IDL) routines. The process followed the standard procedure of NIR spectroscopic data reduction described in previous works \citep[e.g.][]{rogemar2008,diniz15,rogemar2017}.  The K band datacube was already used in \citet{diniz15} and \citet{rogemar2017} to study the molecular gas and stellar kinematics.  The final data cubes were created with a spatial sampling (spaxel) of 0\farcs05 $\times$ 0\farcs05. 

\begin{figure*}
\begin{center}
        \includegraphics[width=0.95\textwidth]{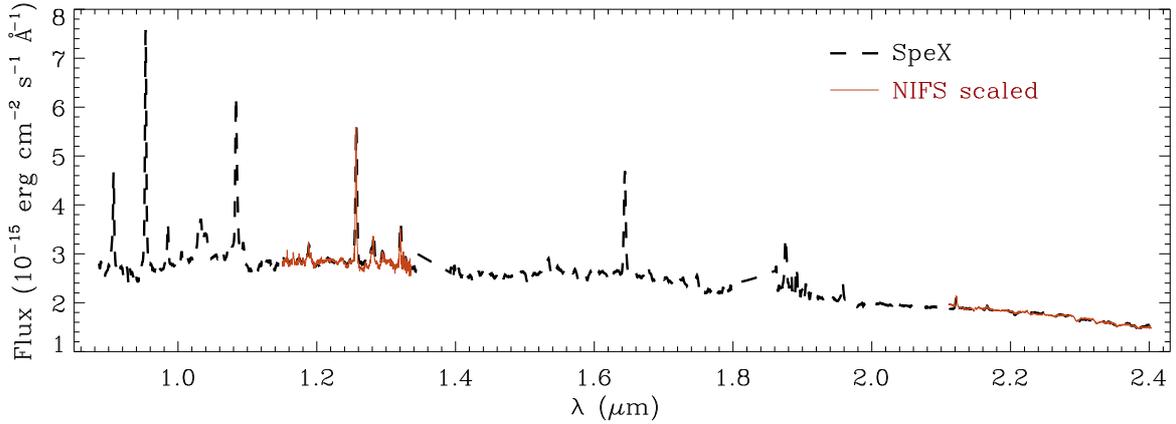}
\caption{\small Comparison of the NIFS (red continuum line) nuclear spectrum after calibration of the datacubes using the IRTF SpeX spectrum of \citet{rogerio06}. Both spectra were obtained with an aperture of 0\farcs8$\times$1\farcs6 oriented along PA$=20^{\circ}$.}
\label{espectro}
\end{center}
\end{figure*}

We have used the long-slit nuclear spectrum of NGC\,2110 presented by \citet{rogerio06} to cross-calibrate the J and K bands in the NIFS data. The long-slit spectrum was obtained with the instrument SpeX of the NASA 3\,m Infrared Telescope Facility (IRTF) within an aperture of 0\farcs8$\times$1\farcs6 with the slit oriented along PA$=20^{\circ}$ using the short cross-dispersed mode that covers the spectral range from 0.8 to 2.4\,$\mu$m. We have extracted the J and K NIFS spectra within the same aperture of the SpeX spectrum, and then scaled them to the same fluxes. In Figure~\ref{espectro} we present the comparison of the final NIFS nuclear spectrum with that of \citet{rogerio06}. Both the continuum and emission-line fluxes are consistent between NIFS and SpeX data. 

The spatial resolution, as obtained from the full width at half maximum (FWHM) of the brightness profile of the telluric standard star, is 0\farcs13 and 0\farcs15 for the J and K band, respectively.  The spectral resolution is $\sim46$~\kms~for both J and K bands, as obtained from the typical FWHM of the arc lamp lines.

\subsection{Spectral synthesis}

The integrated spectrum of an active galaxy consists of several components, such as stellar, gas and dust emission, as well as the AGN torus and accretion disk components. The synthesis process combines a set of spectra of Simple Stellar Populations (SSPs) as well as models for the AGN components, as part of the base of elements used to reproduce the galaxy spectrum. The process allows to retrieve information on the percent contribution of components of different ages and metallicities, as well as of the AGN components. 

We performed spectral synthesis using 
the \starlight~code \citep{asari07,cid04,cid05a,cid05b} in order to quantify the contribution of each component to the spectrum of each spaxel. The code fits the continuum and absorption features of the spectrum by a combination in different proportions of each SSP and AGN components in the base.

We have used the SSP models of \citet{maraston05}, which are described in \citet{rogerio2009}, because they include empirical data with contribution from stars in the TP-AGB phase. This is important because the NIR range presents certain molecular absorption features, such as CN, CO, VO, ZrO and TiO, which are attributed to evolved stars belonging to the Red Giant Branch (RGB) and these TP-AGB stars \citep[e.g.][]{rogerio2015}. 

The SSPs comprise 12 age components ($t=$ 0.01, 0.03, 0.05, 0.1, 0.2, 0.5, 0.7, 1, 2, 5, 9, 13 Gyr) and 4 metallicities (Z = 0.02, 0.5, 1, 2 Z$_\odot$), together with black-body (BB) functions for temperatures in the range $700-1400$ K in steps of 100 K and a power-law ($F_{\nu} \propto \nu^{-1.5}$), in order to account, respectively, for possible contributions from hot dust emission (the AGN torus) and from the AGN.

%Basically, \starlight~solves the following equation for a model spectrum M$_{\lambda}$ \citep{cid05b}:

%\begin{equation}
% M_\lambda=M_{\lambda0}\left[\sum_{j=1}^{N_\star}x_j b_{j,\lambda} r_\lambda \right] \otimes G(\upsilon_\star,\sigma_\star),
%\label{model-spectrum}
%\end{equation}
%where $M_{\lambda0}$ is the synthetic flux at the normalization wavelength free of any emission or absorption line, $\vec{x}$ is the population vector, whose components $x_j$ ($j=1,$...$,N_\star$) represent the fractional contribution of each SSP in the spectral base, $b_{j,\lambda}$ is the normalized spectrum of the $j$th SSP component of the base, $\otimes$ corresponds the convolution operator and $G(\upsilon_\star,\sigma_*)$ is a Gaussian function used to model the line of sight velocity distribution, centered at velocity $\upsilon_\star$ with velocity dispersion $\sigma_*$.  The extinction is modeled assuming a dust uniform distribution in the interestellar medium, following the extinction law of \citet{cardelli89}. 

%The best fits is achieved by minimizing the following equation:

%\begin{equation}
%\chi^2=\sum_{\lambda}[(O_\lambda - M_\lambda)\omega_\lambda]^2,
%\label{minimum}
%\end{equation}
%where $O_\lambda$ is the observed spectrum, $M_\lambda$ is the fitted model,
%$\omega_\lambda = 1/e_\lambda$  and $e_\lambda$ corresponds to the associated uncertainties to the observed spectrum. 

In short, the \starlight~code uses a combination of numeric methods to determine the stellar mass,  extinction ($A_{\rm V}$), ages, metallicities, radial velocity and stellar velocity dispersion . %It seeks to quantify the contribution of each SSP to the observed spectrum of the galaxy.
For each observed spectrum, the synthesis process generates a list of parameters and a synthetic spectrum. % resulting from the best fit to the observed spectrum.
The J- and K-band datacubes were combined using the peak of the continuum emission as reference, and in order to increase the signal-to-noise ratio of the spectra, we have rebined the combined datacube to 0\farcs15$\times$0\farcs15 spaxels before performing the spectral synthesis.

%%%%%%%%%%%%%%%%%%%%%%%%%%%%%%%%%%%%%%%%%%%%%%%%%%%%%%%%55
%%%%%%%%%%%%%%%%%%%%%%%%%%%%%%%%%%%%%%%%%%%%%%%%%%%%%%%%55
%%%%%%%%%%%%%%%%%%%%%%%%%%%%%%%%%%%%%%%%%%%%%%%%%%%%%%%%55

\subsection{Emission-line fitting}

In order to map the emission-line flux distributions, centroid velocities  and velocity dispersion, we have fitted the line profiles with Gaussian curves. The corresponding maps for the K-band emission lines were already presented in \citet{diniz15}; here we focus
 on the J-band. The J-band spectra of NGC\,2110 present strong emission of \pii\ at $1.1886~\mu$m, \feii\ at $1.2570~\mu$m and \pb\ at $1.2822~\mu$m. We have used the  \profit~routine \citep{rogemar2010a} to fit the emission-line profiles. This routine uses the MPFITFUN routine \citep{mark09} to perform a nonlinear least-squares fit and outputs measurements for the flux, centroid velocity and velocity dispersion of each fitted emission-line at each spaxel, as well as their corresponding uncertainties. We do not subtract the stellar population contribution before the fitting of the emission-line profiles, as the SSP models available in the NIR spectral range have much lower spectral resolution than the NIFS data. Indeed, the expected effect on the emission-line fluxes is within the uncertainties \citep{martins13,dametto14}.  The continuum underneath each line is  fitted by a first order polynomial, 
%which results in a good representation of the observed spectra, 
as each emission-line is fitted separately and only a small spectral range is included in the fit.

\begin{figure*}
\begin{center}
        \includegraphics[scale=1]{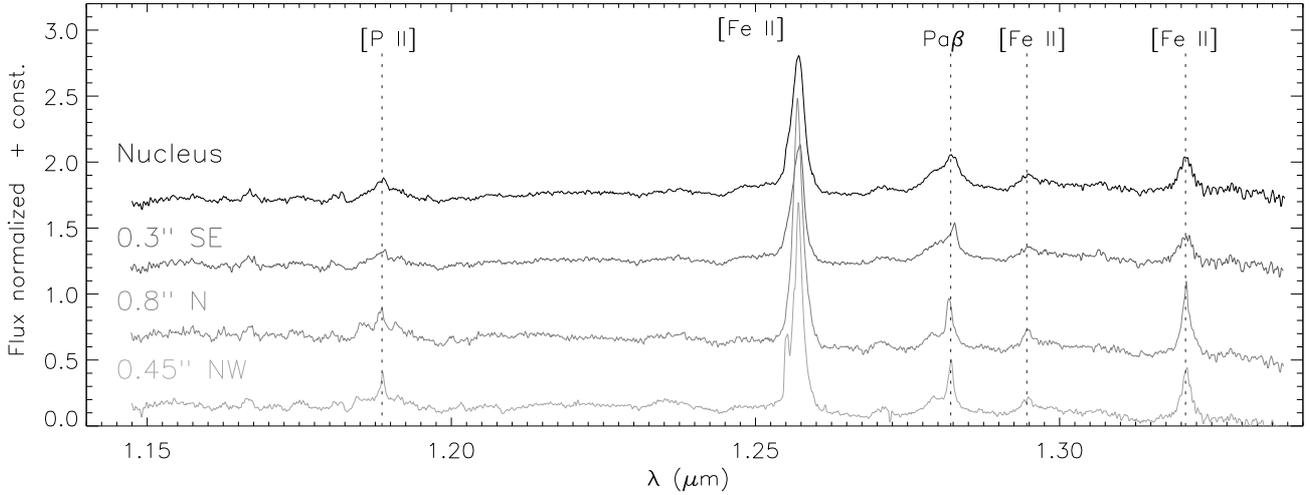}
\caption{\small Sample of spectra in the J-band, in arbitrary flux units per wavelength, corrected for Doppler shift and extracted within a circular aperture of radius 0\farcs25. The locations relative to the nucleus are the same of the spectra shown in Figure 1 of \citet{diniz15}. }
\label{specs}
\end{center}
\end{figure*}

In Figure~\ref{specs} we show examples of the J-band spectra extracted within circular apertures with radius 0\farcs25, centered at the locations of the spectra shown in Figure~1 of \citet{diniz15}. The first spectrum corresponds to the nucleus, the second to a typical extra-nuclear position, the third to a region of enhanced \h2~emission and the last to a location of enhanced \brgamma~and \pb. All spectra present emission lines \pii$\lambda$1.1886\,$\mu$m, \feii$\lambda\lambda\lambda$1.2570,1.2946,1.3209\,$\mu$m and \pb$\lambda$1.2821\,$\mu$m, which can be used to map the kinematics and excitation of the emitting gas.

%%%%%%%%%%%%%%%%%%%%%%%%%%%%%%%%%%%%%%%%%%%%%%%%%%%%%%%%55
%%%%%%%%%%%%%%%%%%%%%%%%%%%%%%%%%%%%%%%%%%%%%%%%%%%%%%%%55
%%%%%%%%%%%%%%%%%%%%%%%%%%%%%%%%%%%%%%%%%%%%%%%%%%%%%%%%55

\section{Results}\label{results}

\subsection{Stellar populations}\label{pop}

\begin{figure*}
\begin{center}
    \includegraphics[scale=1]{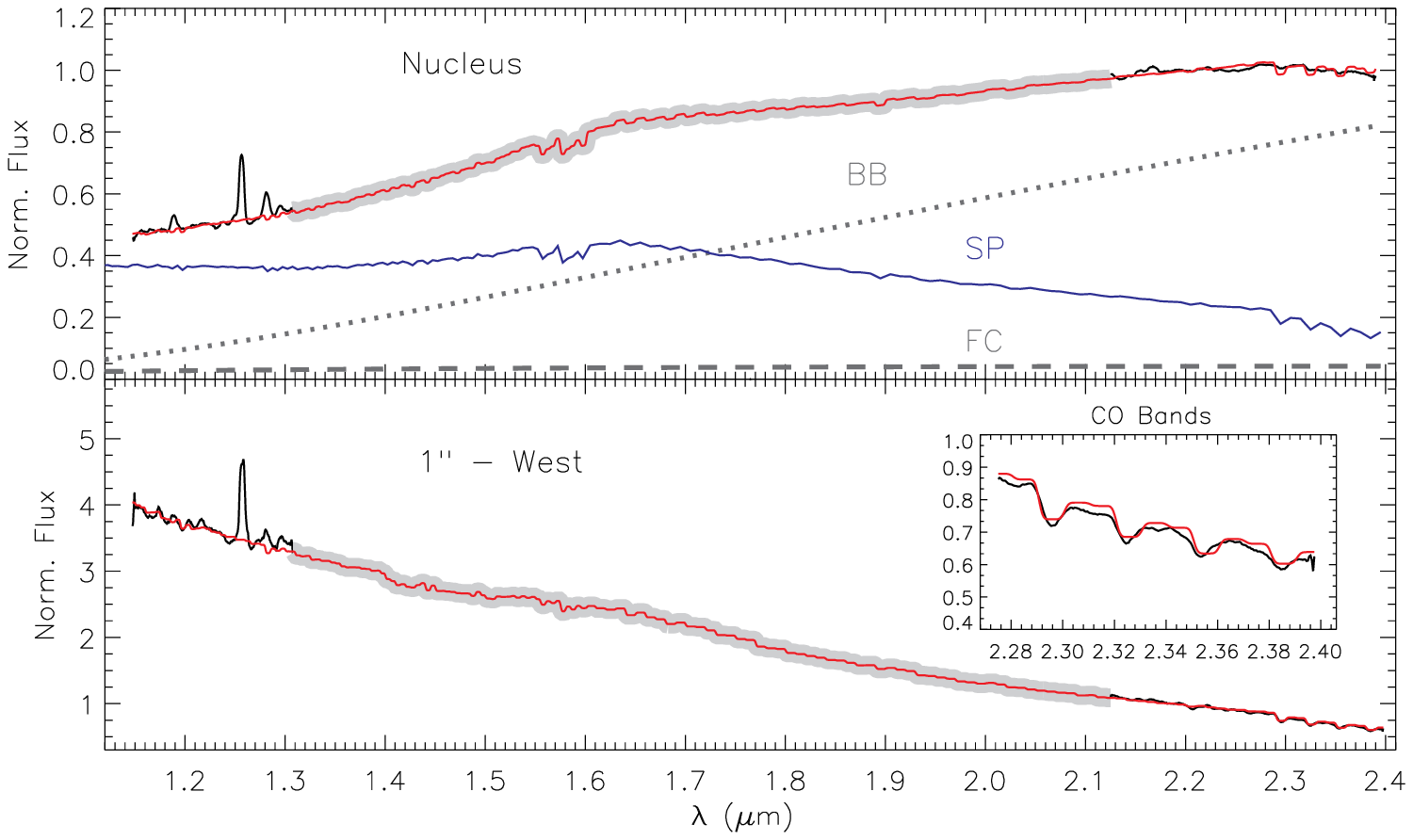} \qquad
    \includegraphics[width=\textwidth]{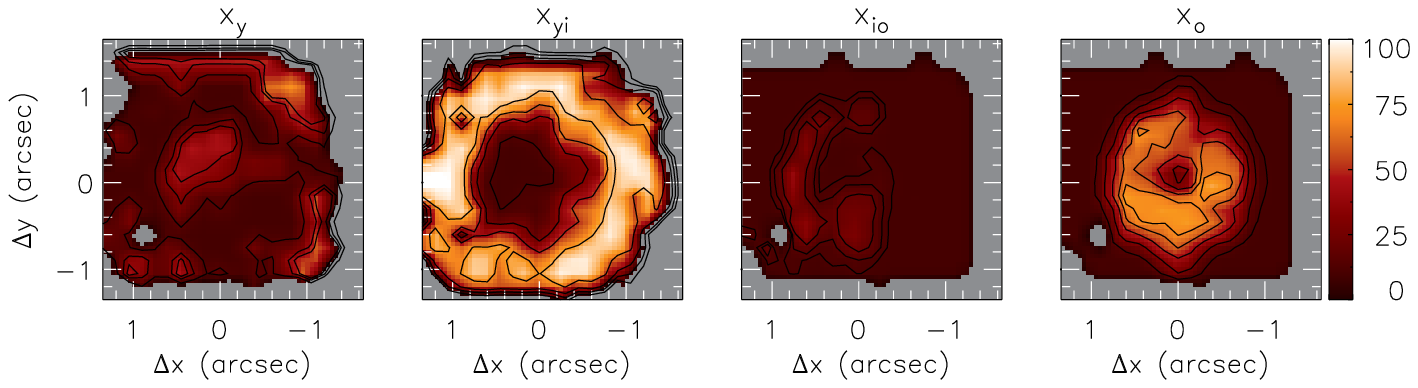} \qquad
    \includegraphics[width=\textwidth]{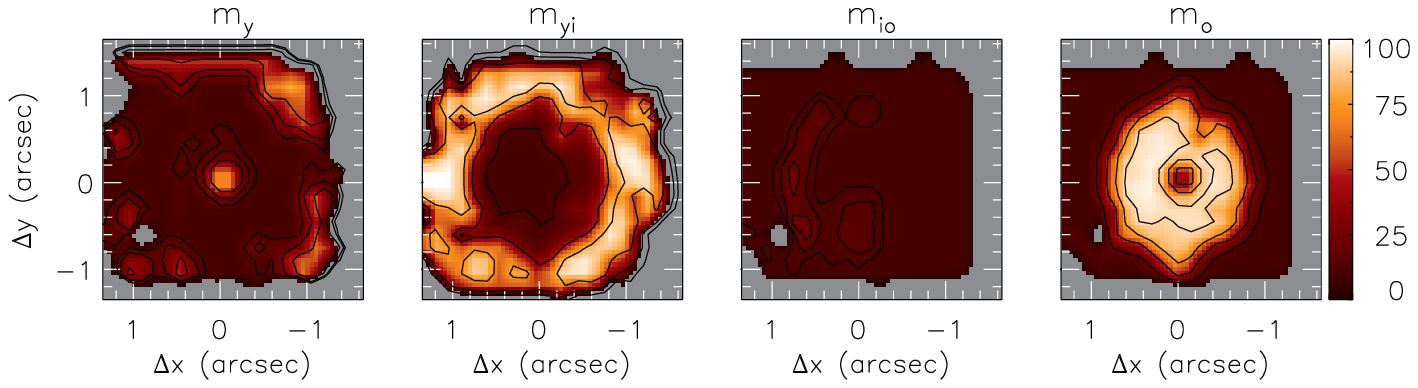}
\caption{\small Top two panels: observed spectra from the nucleus and 1$^{\prime\prime}$ west of the nucleus (in black) compared to the synthesized spectra in red. The gray shaded region corresponds to the H-band spectrum that was not observed. The insert shows a zoom of the CO stellar absorption region. For the nucleus, we show the flux contributions of the SP (25\,\% at 2.2\,$\mu$m), BB (70\,\%) and FC (5\,\%) components separately. Bottom two row of panels, from left to right: in the top row, percent contributions to the 2.2$\mu$m continuum of young (\xy~$\leq 100$\,Myr), young-intermediate ($ 100 < $ \xyi~$\leq 700$\,Myr), intermediate-old ($700\, {\rm Myr} < $ \xio $ \leq 2$\,Gyr) and old ($2 < $ \xo $\leq 13$\,Gyr) age components; in the bottom row, we show the corresponding percent mass contributions (\my, \myi, \mio~and \mo). 
%The spatial orientation of the panels is PA$=133^{\circ}$.
}
\label{bri2110}
\end{center}
\end{figure*}

The top two panels of Figure~\ref{bri2110} display two examples of fits resulting from the spectral synthesis using the \starlight~code. The observed spectra are shown in black, normalized by the flux at 2.1955\,$\mu$m and extracted within apertures of 0\farcs15$\times$0\farcs15. The spectrum of the top panel is centered at the nucleus -- defined as the location of the continuum peak -- and the one of the lower panel is centered at $1''$ west of the nucleus and is representative of typical extra-nuclear spectra. 
The best model synthesized via the \starlight\ code is shown in red.   During the fit of the spectra, emission lines and spurious features (e.g. remaining telluric features or bad pixels) were masked out. The gray shaded region in the spectra of Figure~\ref{bri2110} were not synthesized as we do not have observations in the H band. 

Using the output from \starlight~for each spaxel of the datacube we constructed maps for the percent contributions in flux and stellar mass of each stellar population component found in the galaxy spectra.
Although the base of components has SSP models for 12 different ages, we binned the contributions into only four age components. This procedure is recommended by \citet{cid05b} and has been used previously by our group, since the separation of components with very similar ages and metallicities is not well constrained.
Therefore, the maps of the stellar population components percent contribution, weighted by the flux comprise the following age ranges:
{\bf young} (\xy: $t \le 100\,$Myr); 
{\bf young-intermediate} (\xyi: $100 < t \le 700\,$Myr);
{\bf intermediate-old} (\xio: $700$\,Myr $ < t \le 2\,$Gyr) and
{\bf old} (\xo: $2 < t \le 13\,$Gyr).  Similarly, the mass-weighted ($m_y$, $m_{yi}$, $m_{io}$, $m_o$) component maps were constructed for the same age ranges.

In Figure~\ref{bri2110} we also present the maps for the percent contributions in flux (middle panels) and mass (bottom panels) of the SPs for NGC\,2110. 
We used the mean percent deviation over all fitted pixels, defined as ${\rm Adev}=|O_{\lambda} - M_{\lambda}|/O_{\lambda}$, to exclude regions of poor fits.  $O_{\lambda}$ is the observed spectrum and $M_{\lambda}$ its corresponding best model.  Regions with ${\rm Adev}>12\,$\% were masked out in the maps and are shown in grey.  The cutoff {\rm Adev} value was determined by eye inspection of the resulting fits. The maps for the young-intermediate population clearly show an increase of contributions with the distance of the nucleus, delineating a ring-shaped structure surrounding the nucleus with a radius of $\sim1''$.  Inside the ring the old population (right panels) shows flux and mass contributions of up to 70\% and 100\%, respectively. The intermediate-old population maps do not present significant contributions to the observed fluxes and masses. For the young population a contribution of up to 30\% in flux and mass is observed at the nucleus and in some ``knots"  to the north and northwest of the nucleus at distances $\sim1''$, thus co-spatial with the ring mentioned above.

\begin{figure*}
\begin{center}
    \includegraphics[width=\textwidth]{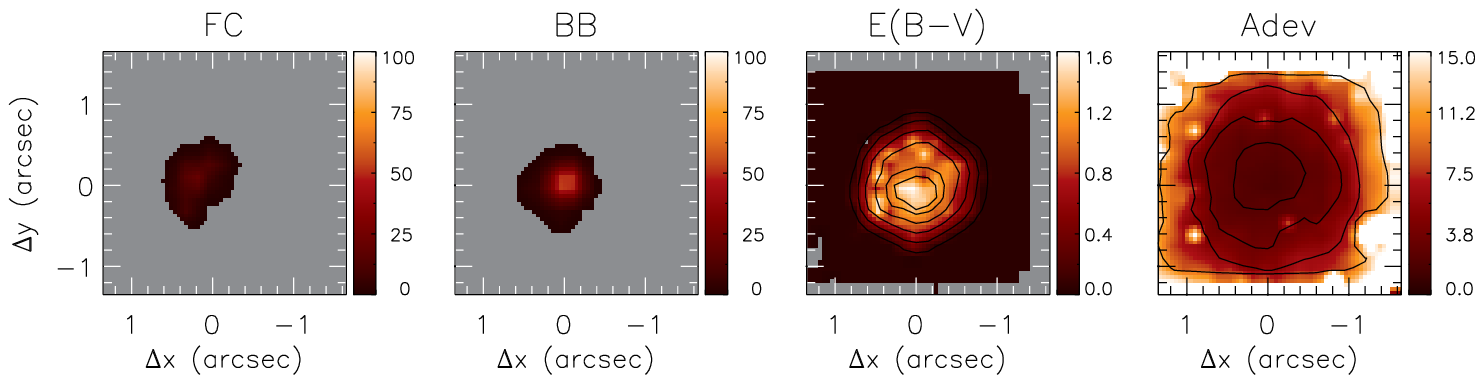}
\caption{\small From left to right: percent  flux contributions to the 2.2$\mu$m continuum of the power-law and black-body components;  reddening $E(B-V)$ derived from the stellar population synthesis and the parameter ${\rm Adev}$, which quantifies the quality of the fit.}
\label{fc2110}
\end{center}
\end{figure*}

Besides the flux and mass contributions of the SPs, the \starlight\ code outputs also the contributions of the featureless continuum (FC) and BB components to the light at 2.2$\mu$m.
The corresponding maps of these components are shown in Figure~\ref{fc2110}, together with maps for the color excess [$E(B-V)$] and ${\rm Adev}$. In order to construct the map for the BB percent contribution, we have summed the contributions of all black-body functions in the base. The FC and BB maps show a nuclear unresolved structure, with contributions of up to 10\% and 35\% , respectively. Such unresolved structure is expected in AGN considering the spatial resolution of the NIFS data (0\farcs15 or 20 pc), the FC due to continuum emission from the accretion disk and the BB due to emission from the putative dusty torus. The ${\rm Adev}$ parameter -- that measures the quality of the fit -- present values smaller than  7\,\% at most locations, indicating the resulting fits reproduce well most of the observed spectra.

\subsection{Emission-line flux distribution and gas kinematics}\label{resul_flux_kinem}

\begin{figure*}
\begin{center}
    \includegraphics[width=0.99\textwidth]{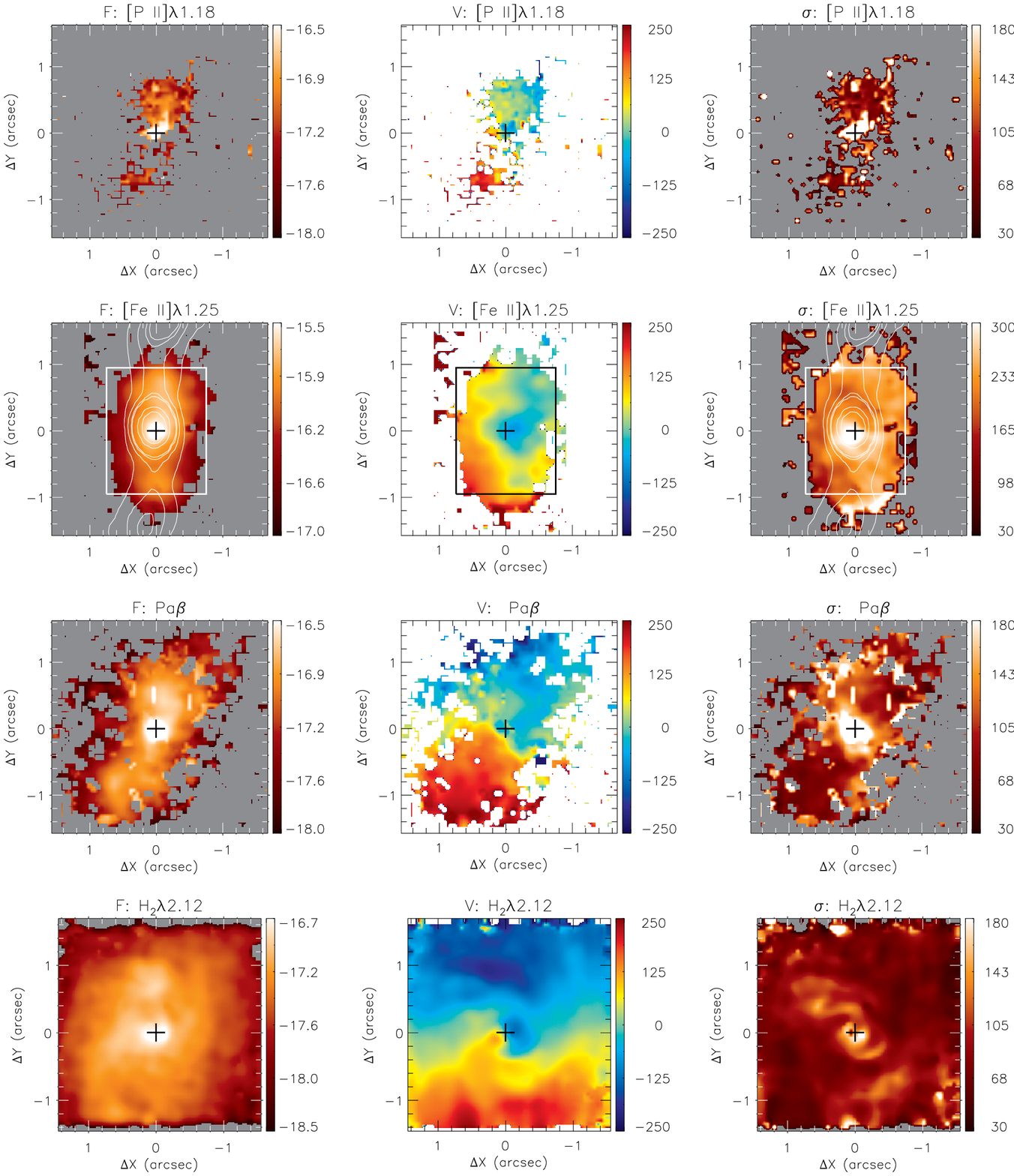}
\caption{Flux distributions (left panels), velocity fields (middle panels) and velocity dispersion maps (right panels) for the [P\,{\sc ii}]$\lambda1.18\,\mu$m, [Fe\,{\sc ii}]$\lambda1.25\,\mu$m, \pb\ and H$_2\lambda2.12\,\mu$m emission lines. The white contours overlaid to the \feii\ flux and $\sigma$ maps are from the 6\,cm radio image of \citet{ulvestad84}. White rectangles shown in the \feii\ maps indicate the FOV of the observations used by \citet{durre14}. The ``+" sign marks the location of the nucleus.}
\label{gas}
\end{center}
\end{figure*}

Figure~\ref{gas} presents the flux, centroid velocity and velocity dispersion for the following emission lines: \pii$\lambda1.1886\,\mu$m, \feii$\lambda1.2570\,\mu$m, \pb~and \h2$\lambda2.1218\,\mu$m. These are the strongest NIR emission lines present in the spectra of the nuclear region of NGC\,2110. Although \citet{diniz15} have already presented the maps for the K-band emission lines, we present again the \h2$\lambda2.1218\,\mu$m maps to allow a comparison of the emission distribution and kinematics of the ionized and molecular gas.  We have excluded regions where the measurements have uncertainties larger than 50\% in flux and larger than 50\kms\ in the centroid velocity or velocity dispersion. The masked regions are shown in gray in the flux and velocity dispersion maps, and in white in the velocity maps.

The flux distributions of all lines peak at the nucleus, adopted as the location of the continuum peak emission, indicated by a cross in the maps. In order  to evidence also the emission of lower flux structures,  we have used  a logarithmic  scale  in the  maps. These maps show that the flux distributions of distinct emission lines are clearly different from each other.  The molecular gas emission is observed over the whole field of view (FOV) presenting two regions of strongest emission, one at the nucleus and other at $\sim$0\farcs7 north of it. \pb~shows extended emission mostly along the southeast--northwest direction, similarly to the Br$\gamma$ flux distribution presented in \citet{diniz15}. The \feii\ flux map presents also a collimated structure, but oriented along the north--south direction. The \feii\ emission seems to follow the direction of the radio jet, as indicated by the white contours overlaid to the \feii\ flux and $\sigma$ maps. The radio emission contours are from the 6\,cm image of \citet{ulvestad84}, obtained with the Very Large Array (VLA). The white rectangles shown in the \feii\ flux and $\sigma$ maps delineate the FOV of the Keck OSIRIS observations used by \citet{durre14}, which will be discussed in Sec.~\ref{discussion}.
\pii\ shows a similar flux distribution to that \pb, although weaker thus apparently less extended.

The centroid velocity maps are shown in the middle row of panels of Figure~\ref{gas}. The color bars show the velocity values in units of \kms, relative to the systemic velocity ($2332$~\kms, corrected to the heliocentric rest frame),  which was determined from the  fit of  the stellar velocity  field (for more details see \citet{diniz15}).  The H$_2$ velocity field was already discussed in \citet{diniz15} and suggest the presence of three components: (i) a rotation pattern with blueshifts to the north and redshifts to the south and a velocity amplitude of $\sim250$\kms; this rotation pattern is similar to that seen for the stars \citep{diniz15,rogemar2017}; (ii) gas inflows along nuclear spirals to the north and south of the nucleus and (iii) a compact biconical outflow along the east--west direction, seen within the inner 0\farcs5.  

The \pb\ velocity field is similar to that of Br$\gamma$ shown in \citet{diniz15}, but as the line is stronger we were able to fit it over a larger region than for Br$\gamma$. As for the H$_2$, the \pb\ velocity field shows blueshifts to the northwest and redshifts to the southeast of the nucleus, which is consistent with rotation in the plane of the galaxy. However, for \pb\ the components (ii) and (iii) reported above are not seen.
 The rotation component seems also to be present in the \feii\ velocity field, but this line kinematics shows additional components, such as the blueshifts at the nucleus and redshifts to the east and northeast of the nucleus.

The bottom panels of Figure~\ref{gas} show the $\sigma$ maps. For \h2, the highest values ($\sim$150\kms) are observed at the nucleus and along PA$\sim$45$^{\circ}$/225$^{\circ}$, adjacent from the location of the ``kinematic spiral" (systematic residuals relative to a rotation pattern following the shape of spiral arms) seen in the H$_2$ velocity field \citep{diniz15}. The \pb\ shows the highest $\sigma$ values ($\sim$180\,km\,s$^{-1}$) at the nucleus, while in the extra-nuclear region the values are similar to those of H$_2$. Although the \pii~$\sigma$ map is noisier, as pointed out above, the $\sigma$ values are similar to those of \pb.  The \feii\ $\sigma$ map presents overall higher values, of up to 300~\kms, with the highest ones seen at the nucleus, while at most locations $\sigma>150$\,\kms.

\subsection{Emission-line ratios}\label{ratios}

\begin{figure*}
\begin{center}
    \includegraphics[width=\textwidth]{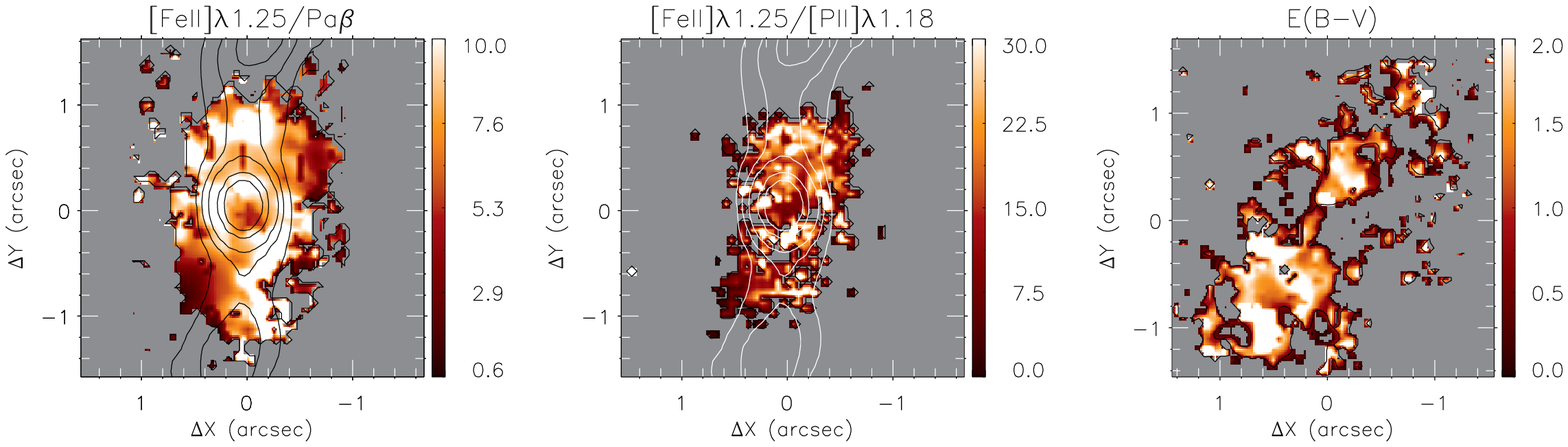}
\caption{\small \feii$\lambda1.25\mu$m/\pb~line ratio map (left panel), \feii$\lambda1.25\mu$m/\pii$\lambda1.18\mu$m line ratio map (center panel) and reddening map obtained from the \pb/\brgamma~line ratio (right panel). The contours shown in the first two maps trace the radio emission.}
\label{raz}
\end{center}
\end{figure*}

In order to investigate the excitation mechanisms of the \feii\ emission lines, we constructed the line ratio maps shown in Figure~\ref{raz}. In each panel, grey regions correspond to masked locations where we were not able to fit one or both lines of the corresponding ratio map, following the same rejection criteria described above.  %The green contours are from the 6\,cm radio image of \citet{ulvestad84}. 

For \feiiumdois/\pb, the values range from 1 to 10. The highest ones seem to be correlated with the radio contours, while the lowest ones are seen at $\sim$0\farcs7 northwest and $\sim$0\farcs8 southeast of the nucleus. At the nucleus this ratio is $\approx6$.

Although the \feiiumdois/\piium~map is noisy, values of up to 30 can be observed, with the highest ones surrounding the radio structures and the lowest ones ($\approx7$) at similar locations where the \feiiumdois/\pb~ratio map also presents its lowest ratios.

The $E(B-V)$ map shown in the right panel of Figure~\ref{raz} was constructed using the \pb/Br$\gamma$ emission-line ratio, via the following equation:
\begin{equation}
 E(B-V)=4.74\,{\rm log}\left(\frac{5.88}{F_{Pa\beta}/F_{Br\gamma}}\right),
\end{equation}
where $F_{Pa\beta}$ and $F_{Br\gamma}$ are the fluxes of Pa$\beta$ and Br$\gamma$ emission lines, respectively. We used the reddening  law of \citet{cardelli89} and adopted the intrinsic ratio between \pb\ and Br$\gamma$ of $F_{Pa\beta}/F_{Br\gamma}=5.88$, corresponding to case B at the low-density limit \citep{osterbrock06} for an electron temperature of $T_e=10^4$\,K.  As shown in \citet{diniz15}, the Br$\gamma$ emission is detected only in a strip along the northwest--southeast direction. Thus, we map the reddening only at these locations. The corresponding $E(B-V)$ map for NGC\,2110 displays the highest values of 2 mag, while the mean $E(B-V)$ value is $\approx1.4$\,mag.

\section{Discussion}\label{discussion}

\subsection{Distribution of the stellar populations}\label{discupop}

The distribution of the stellar populations of NGC\,2110 (Figure~\ref{bri2110}) indicates the old population contributing most to the observed continuum emission within the inner $\approx$0\farcs8 ($\approx$120\,pc), while the intermediate-age stellar population shows a ring-shaped structure at larger distances from the nucleus. The young stellar population presents the highest flux contributions at the nucleus and close to the borders of the FOV, in the external borders of the ring seen in the intermediate age component. 

The presence of an old stellar population around the nucleus of NGC\,2110 is in good agreement with  previous  NIR \citep{rogerio2009} and optical  \citep{delgado2001,raimann03} studies. This is also consistent with  results found by our group for others Seyfert galaxies -- Mrk\,1066 \citep{rogemar2010b}, Mrk\,1157 \citep{rogerio2011c}, NGC\,1068 \citep{ngc1068} and Mrk\,573 \citep{diniz17} -- using NIR integral field spectroscopy. 

\begin{figure}
\begin{center}
    \includegraphics[scale=0.65]{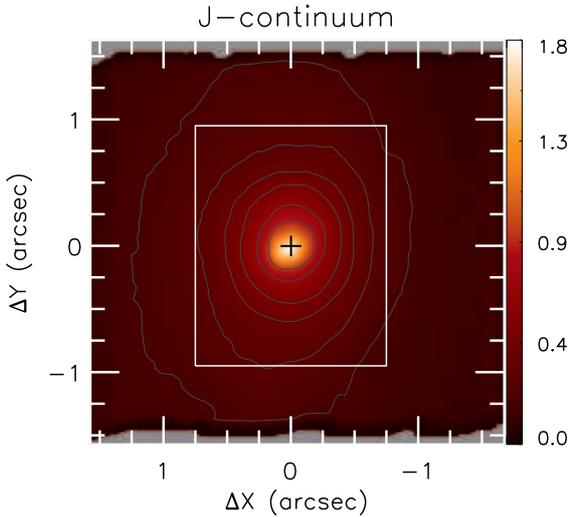}
\caption{\small J-band  continuum image obtained by averaging the NIFS datacube in the spectral range 1.228--1.289~$\mu$m. The color bar shows the fluxes in units of  $10^{-17}$\,erg\,s$^{-1}$\,cm$^{-2}$\,\AA$^{-1}$. The white rectangle corresponds to the FOV of the observations used by \citet{durre14}.}
\label{contj}
\end{center}
\end{figure}

The flux contributions of the young stellar population  in the central region of NGC\,2110 are co-spatial with locations where the H$_2$~emission is more intense (Figure~\ref{gas}). 
A similar correlation between the young population and the molecular gas distribution was previously observed for  NGC\,1068 \citep{ngc1068}, indicating that the molecular gas represents the circumnuclear gas reservoir required for star formation.
%The [Fe {\sc ii}]$\lambda$1.2570\,$\mu$m flux map (Figure~\ref{gas}) does not show any signature of enhanced emission at the locations where star clusters were detected by \citet{durre14}.
Based on the J-band continuum image, \citet{durre14} detected four star clusters with estimated ages of 1-3~Myr, surrounding the nucleus of NGC\,2110 at distances smaller than 0\farcs2.
Our J-band continuum (Figure~\ref{contj} emission image constructed from the NIFS datacube in the same spectral range (1.228-1.289~$\mu$m) used by them, does not show any clear knot of emission associated to the unresolved regions seen in their continuum image. The discrepancy between OSIRIS and NIFS continuum images is possible due to a slightly better angular resolution of the OSIRIS data.
On the other hand, a young stellar population component is found in our spectral synthesis (Figure~\ref{bri2110}), but it should be noted that our spectral base does not include SSP components younger than 10~Myr. Furthermore,  reddened young stars of age of a few million years are  indistinguishable from an FC component. 

The 150\,pc radius ring-shaped structure observed in the young-intermediate stellar population component both in flux and mass seems to be weakly correlated with the lowest stellar velocity dispersion ($\sigma_\star$) values presented in \citet{diniz15} and \citet{rogemar2017}. Similar rings of young/intermediate-age stellar populations associated to low $\sigma_\star$ have been previously reported by our group for other Seyfert galaxies  \citep{rogemar2010b,rogerio2011c,diniz17,natacha19}, suggesting that this may be a frequent characteristic of the central region of AGN. A possible scenario for this is: if there is plenty of gas moving towards the central region of the galaxy, this gas may find suitable conditions for the formation of new stars in the inner few 100\,pc. Part of this gas will continue falling towards the center, and/or the mass loss from the recently formed massive stars will move inwards triggering and feeding an AGN there. Low $\sigma_\star$ rings, seen in the central region of AGNs would thus be originated in intermediate-age stars, which have formed from low-velocity dispersion gas, tracing the cold kinematics of the gas cloud that formed them \citep{rogemar2017}. 

Flux and mass contribution maps of each stellar population component (Figure~\ref{bri2110}) show that the old population contributes most inside the young-intermediate age ring, while the young stellar component is found within $\approx$0\farcs5 ($\sim$70\,pc) from the nucleus and also at the outer borders of the ring.

\subsection{Extinction}

The average reddening value for the gas (right panel of Figure~\ref{raz}) is $E(B-V)\sim1.4$ mag, while for the stars (third panel of Figure~\ref{fc2110}) $E(B-V)\sim0.8$ mag. In fact, a larger reddening is expected for the gas than for the stars, since the former should be concentrated in the galactic disk, while the stars reach higher latitudes in the galaxy bulge, where the reddening is lower \citep{calzetti94}.

Our $E(B-V)$ map for the gas is in good agreement with previous estimates  using the \feii~$\lambda$1257/$\lambda$1644 flux ratio \citep{durre14} and  using the \pb\ and \brgamma\ fluxes \citep{thaisa99} from long-slit spectra.

The stellar $E(B-V)$ values obtained by us are larger than those  obtained by \citet{raimann03} ($\sim0.4$ mag) using  optical observations, but they are consistent with those reported by \citet{rogerio2009} ($\sim0.6$ mag) from NIR data. Higher reddening values determined using NIR data as compared to those using optical determinations can be interpreted as due to the fact that NIR observations are less affected by dust extinction and thus probe more obscured regions immersed in the plane of the galaxy \citep{martins13,rogemar15}.

\subsection{Hot dust emission and AGN Featureless continuum}

Our synthesis results detected a small contribution (10\,\%) of the power-law FC component, which is associated with the radiation from the accretion disk. Although the FC component is expected to be more important at shorter wavelengths, it is usually necessary to properly fit the NIR spectral energy distribution as well. \citet{rogerio2009} presented spectral synthesis of the nuclear NIR spectra of a sample of 24 Seyfert galaxies and found that at least 50\% of the Seyfert\,2 and 60\% of the Seyfert\,1 require a power-law component to properly fit their spectra.  
For NGC\,2110, they found a contribution of FC=33.6\% to the flux at 1.2\,$\mu$m, which translates to about 13\% at 2.2~$\mu$m (our normalization point). Thus, our result is in good agreement with those of \citet{rogerio2009}.

We found that the hot dust emission, traced by the BB component, contributes with 35\% of the 2.2~$\mu$m unresolved nuclear continuum emission. This value is larger than that found by  \citet{rogerio2009} of only 1\% at 1.2~$\mu$m. %However, it should be noted  that the contribution of hot dust emission is expected to be larger to the K-band than to the J-band continuum.  
In addition, the spatial resolution of our data is about one order of magnitude better than those of \citet{rogerio2009}, which allow us to better probe the unresolved dust emission. Similar discrepancies of the nuclear contributions of BB components between high and low resolution data have already been reported for other objects \citep[e.g.][]{rogemar2010b}. To determine the average dust temperature, we  fitted an integrated spectrum extracted within an 0\farcs3 radius aperture centred at the peak of the continuum emission. Three dust components were used by {\sc starlight} in the fit (with temperatures of 1200, 1300 and 1400~K) and the resulting dust temperature is $\sim$1340\,K, obtained by computing an average temperature weighted by the integrated flux contribution of each component.  

% and is going through the torus.
%The detection of this component supports the results obtained by \citet{veron2006}, where they detected a reddened \pb~broad-line region   

From the contributions of the BB components to the observed continuum we can estimate the mass of the hot dust that originates these components. The total mass of hot dust ($M_{\rm HD}$) at the nucleus of NGC\,2110 was estimated following \citet{rogerio2009}, using the formalism of \citet{barva1987} and assuming the dust composed by grains of graphite.

The IR spectral luminosity of each dust grain, in erg\,s$^{-1}$\,Hz$^{-1}$, 
can be written as 
\begin{equation}\label{lumigrain}
 L^{\rm gr}_{\nu,{\rm ir}} = 4\pi^2 a^2 Q_\nu B_\nu(T_{\rm gr}),
\end{equation}
where $a = 0.05\,\mu$m is the grain radius, $Q_\nu = 1.4 \times 10^{-24}\nu^{1.6}$
is its absorption efficiency and
$B_\nu(T_{\rm gr})$ is its spectral distribution assumed to be a Planck function
for a temperature $T_{\rm gr}$.

The total number of graphite grains can be obtained from
\begin{equation}
 N_{\rm HD} \sim \frac{L^{\rm HD}_{\rm ir}}{L^{\rm gr}_{\rm ir}},
\end{equation}
where $L^{\rm HD}_{\rm ir}$ is the total luminosity of the hot dust,
obtained by integrating the flux of each BB component contribution from the synthesis. Then, we multiplied the integrated normalized flux by the normalization flux at 2.1955\,$\mu$m, and converted it to the correct units.
%(from erg\,s$^{-1}$\,cm$^{-2}$\,\AA$^{-1}$ to erg\,s$^{-1}$\,Hz$^{-1}$). 
In order to obtain $L^{\rm gr}_{\rm ir}$, we have integrated the Eq.~\ref{lumigrain}
for all temperatures, ranging them from 700 to 1400\,K, in steps of 100\,K.

Under these assumptions, $M_{\rm HD}$ can be obtained by \citep[e.g.][]{ardila2005}:

\begin{equation}
 M_{\rm HD} \sim \frac{4\pi}{3}a^3N_{\rm HD}\rho_{\rm gr},
\end{equation}
where $\rho_{\rm gr} = 2.26$\,g\,cm$^{-3}$ is the density of the grain. 
The total dust mass estimated for the nucleus of NGC\,2110 
%by integrating over the entire field of view, 
%***[COMENTEI "OVER THE ENTIRE FOV"  PORQUE ISTO E PARA SER NUCLEAR, E A EMISSAO SO SE OBSERVA MESMO BEM PROXIMA AO NUCLEO!]*** 
is $M_{\rm HD}\sim3\times10^{-2}$\,M$_{\odot}$, which is within the range of masses obtained for other active galaxies ($10^{-4}-10^{-1}$\msun) \citep[e.g.][]{ardila2005,ardila2006,rogerio2009,rogemar2009a,diniz18}.

\subsection{Gas excitation}

A widely studied issue in the NIR is how to determine the main mechanism that originates the lines of \feii~and \h2. In \citet{diniz15} we conclude that the \h2~emission is due to thermal processes (heating of the gas by X-rays from the AGN and/or shocks), with an excitation temperature in the range $\sim2100-2700$\,K. Here we discuss the origin of the \feii\ emission.

The \feii~emission lines can be produced by collisional excitation due to the heating of the gas by shocks \citep{hollenbach89} or by X-rays from the AGN \citep{maloney1996}. The key point is to determine whether the excitation of \feii~is due to X-rays or shocks.  
As pointed out in Sec.~\ref{resul_flux_kinem}, the kinematics of the \feii~emitting gas reveals a distorted velocity field, higher $\sigma$ values and the flux distribution correlated with the radio jet, 
supporting the contribution of shocks to the \feii~emission (see Sec.~\ref{disc-kin} for more details on the interpretation of the gas kinematics).

One way to investigate the origin of \feii~is using emission-line ratios, such as \feiiumdois/\pb~and \feiiumdois/\piium. The \feii/\pb~line ratio presents values between $0.6$ and $2$ for Seyfert nuclei, lower than $0.6$ for Starburts, and higher than $2$ for Low-ionization nuclear emission-line regions (LINERs) \citep{reunanen02,ardila2004,ardila2005b,rogerio2013,colina15}. The \feii/\pb~ratio map (left panel of Figure~\ref{raz}) presents values higher than $2$ for most of the observed field, with the highest ones being co-spatial or surrounding the radio structure. 
This suggests that the radio jet plays an important role in the origin of the \feii\ emission.

The conclusion above is also supported by the \feiiumdois/\piium~map shown in the central panel of Figure~\ref{raz}. These lines have similar ionization temperatures, with their parent ions presenting similar ionization potentials and radiative recombination coefficients. In \hii~regions, where shocks are not important, the \feii/\pii~ratio is $\sim2$. For supernova remnants, where shocks are the dominant excitation mechanism, line ratios are typically larger than 20 \citep{oliva2001}. In this case, fast shocks have passed through the gas destroying the dust grains, releasing the Fe and thus enhancing its abundance and its emission. The same can happen in the central region of active galaxies, where nuclear jets can produce shocks \citep[e.g.][]{oliva2001,thaisa2009}. For NGC\,2110,  \feii/\pii~values reach up to $30$, being observed at the same locations of the highest \feii/\pb~ratios, supporting the contribution of shocks to the emission of the \feii~line.  Thus, based on the observed emission-line ratios, as well as on the correlation among the $\sigma$ values, flux distribution and the radio structure, we conclude that the main excitation mechanism of \feii\ in NGC\,2110 are shocks produced by the interaction of the radio jet with the ambient gas.

\citet{rosario2010} have concluded that the radio jet escapes the disk towards high latitudes and a direct interaction of the gas near the plane of the galaxy with the radio jet would be unlikely. However, the spatial resolution of our data ($\sim$30\,pc) is better than theirs, resolving structures smaller than the disk scale height. We have thus apparently been able to probe this interaction via the \feii\ emitting line. 
%***[SO PARA ESCLARECER: NAO É SO O FE QUE E EJETADO, TODO O GAS E EJETADO; O QUE ACONTECE E QUE A EMISSAO DO FE E UM TRACADOR DO PROCESSO; DO JEITO QUE ESTAVA ESCRITO PARECIA QUE SO O FE QUE ERA EJETADO]***.
Our proposed scenario is thus that the radio jet is pushing away gas from the disk of the galaxy in the vicinity of the AGN, launching it to larger latitudes above the plane of the galaxy.

\subsection{Gaseous Kinematics}\label{disc-kin}

Previous studies of NGC\,2110 at optical and NIR wavelengths have shown a complex gas kinematics in the central region of the galaxy. Asymmetries in the rotation curve of the galaxy and distinct kinematic components have been observed \citep[e.g.][]{thaisa99,knop2001,delgado2002,ferruit2004,allan2014,diniz15}. In \citet{diniz15} we found that the H$_2$~emitting gas shows three distinct kinematic components: gas rotation in the plane of the disk, streaming motions towards the center along nuclear spiral arms and a compact outflow within the inner 70~pc oriented along the east--west projected direction. The disk component and the compact outflow are also observed using optical IFS \citep{allan2014}, which reveals also a warm -- higher velocity dispersion -- disk and a high-latitude cloud observed in blueshift at 1\farcs5--4$^{\prime\prime}$ north of the nucleus.

Although all velocity fields of the emitting gas (Figure~\ref{gas}) show an overall similar rotation pattern, with blueshifts to the northwest and redshifts to the southeast, it can be clearly seen that distinct emission lines present somewhat distinct kinematics. The $\sigma$ maps show that at most locations the \h2~presents the lowest $\sigma$ values, followed by \pb~and \pii, with \feii~presenting the highest values, which seem to be correlated with the radio emission.
The velocity field of the molecular gas -- already discussed in \citet{diniz15} -- is completely distinct from that of the ionized gas, and differences between the velocity fields for the  \feii\ and H recombination lines are also clearly present.

The \pb~kinematics is similar to that of \brgamma~from \citet{diniz15} and consistent with the presence of the warm gas disk component detected by \citet{allan2014}. The velocity fields are dominated by rotation in the galaxy disk and their emission can be attributed  as being due to the heating and ionization of the gas by the AGN radiation.

Comparing our \feiiumdois~kinematic maps with those obtained by \citet{durre14} for the same emission line, we find that they are consistent each to other, considering that the spatial orientation of the maps originally published is incorrect. Their maps are rotated by 90$^\circ$, to have the north-south direction along the y-axis (Durr\'e, M., private communication). Considering this, it can be noticed that both their and our $\sigma$ maps show a correlation with the radio structure, which is clearly observed as enhanced $\sigma$ values surrounding the radio structure.   
%Comparing our \feiiumdois~kinematic maps with those obtained by \citet{durre14} for the same emission line, we can see some differences. 
%Their velocity dispersion map does not show a correlation with the radio structure, which is clearly observed as enhanced $\sigma$ values surrounding the radio structure in our measurements (Figure~\ref{gas}).
It is worth noticing that our \feii~kinematics is similar to that of the \nii~nuclear component detected by \citet{allan2014}, although the spatial resolution of our data is about 5 times better than theirs. 
%As mentioned in Sec.~\ref{discupop}, in the \feiiumdois~flux map of \citet{durre14} the highest values are seen in an elongated structure of  $90\times35$\,pc$^2$ with an orientation tilted by $45^{\circ}$ relative to the radio axis, while our \feiiumdois~flux map clearly shows an orientation along the radio jet. 
Recently, \citet{rosario19} presented the Atacama Large Millimeter Array (ALMA), Hubble and Spitzer Space Telescopes, and the Very Large Telescope/SINFONI observations of NGC\,2110 and found that the cold molecular gas presents a lacuna co-spatial with the highest emission of ionized and hot molecular gas, indicating that the AGN is directly influencing the properties of the cold molecular gas. 
In summary, we have found that: (i) the \feii~emission is oriented along the radio jet; (ii) its velocity dispersion map shows enhanced values surrounding the radio structures; (iii) the emission-line ratios support the presence of shocks and (iv) the \feii\ velocity field shows distortions relative to a rotating disk. Considering all this, we suggest that besides rotation, the \feii\ kinematics supports the presence of an additional outflowing component due to the interaction of the radio jet with the ambient gas. The interaction of the radio jet with the ambient gas is supported by its observed orientation \citep[which is displaced by 29$^\circ$ from the major axis of the galaxy and inclined by 40$^\circ$ relative to the plane of the sky ,][]{pringle99}, supporting an interception with the disk of the galaxy for typical disk scale heights. This is also consistent with the CO emission distribution presented by \citet{rosario19} and similar results for the gas kinematics have been observed also for other Seyfert galaxies \citep[e.g.][]{ferruit99,rosario2010,rogemar2006,rogemar15b}.

\subsubsection{The mass outflow rate}

\begin{figure*}
\begin{center}
    \includegraphics[scale=0.7]{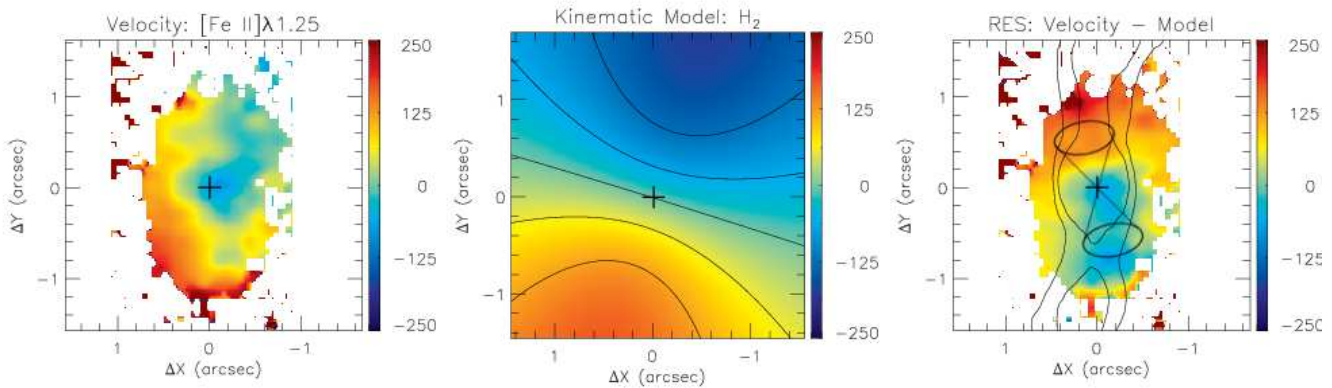}
\caption{\small Observed velocity field for the \feii\ (left), rotating disk model for the H$_2$ from \citet{diniz15} (middle) and residual map obtained by the subtraction of the model from the observed \feii\ velocity field (right). The conical shape drawn at the right panel represents the adopted geometry used to estimate the mass-outflow rate.}
\label{res}
\end{center}
\end{figure*}

Aiming to isolate the outflowing gas component in the \feii\ kinematics, we have used the rotating disk model for the molecular gas (which is dominated by rotation) presented in \citet{diniz15} to construct a residual map between the observed \feii\ kinematics and this model under the assumption that the rotation component should be similar for both gas species. Figure~\ref{res} presents the \feiiumdois~velocity field (left panel), the rotating disk model for the H$_2$ (middle panel) and the residual map (right panel). Contours from the radio image are overlaid on the residual map. 

The residual map clearly shows excess blueshifts at the nucleus and to the south and redshifts to the north, with absolute values of up to 200\,\kms, mainly oriented along the radio structure. These residuals, together with the correlation of the \feiiumdois\ fluxes, $\sigma$ values and line ratios with the radio structure lead us to interpret that the residuals are tracing a biconical outflow possibly originated from interaction of the radio jet with the \feii\ emitting gas.  
This conclusion is supported by previous studies of other similar objects, which have shown  the \feii~has a more disturbed kinematics than the molecular and atomic hydrogen, being considered a tracer of the AGN feedback  \citep[e.g.][]{hicks2009, sanches2009, rosario2010, thaisa2010, rogemar15b}.

%\subsection{Mass outflow rate}

We estimate the ionized gas mass outflow rate through a circular cross-section with radius $r=$0\farcs25 ($\sim36$\,pc)  at a distance of $h=$0\farcs5 ($\sim73$\,pc) from the nucleus, using
%***[NA FRASE  QUE SEGUE: ONDE USA O OPENING ANGLE DO CONE? ESTA GEOMETRIA PODERIA SER USADA PARA DETERMINAR O VALOR DO FILLING FACTOR! NAO VEJO ISTO ABAIXO, SERIA MELHOR USAR MESMO PARA DETERMINAR f! SE NAO USAR, NAO TEM SENTIDO ASSUMIR ISTO!]Assuming a biconical outflow with an opening angle $\alpha=60^{\circ}$ (estimated ***from*** Figure~\ref{res}), we obtain the mass outflow rate using
\begin{equation}
\dot{M}_{\rm out} = 2\,m_{\rm p}\,N_e\,v_{\rm out}\,f\,\,\pi\,r^2,
\label{mout}
\end{equation}
\noindent where $m_{\rm p}$ is the proton mass, $N_e$ is the electron density, $v_{\rm out}$ is the velocity of the outflowing gas ($v_{\rm out}=v_{\rm obs}/\sin\,\theta$), $f$ is the filling factor and $\theta$ is the angle between the axis of the cone and the plane of sky. We have assumed $N_e=500$\,cm$^{-3}$ which is a typical value for outflowing regions around Seyfert nuclei as derived using long-slit spectroscopy \citep[e.g.][]{dors14}, $f=0.01$ (which is a value commonly used for other Seyfert galaxies -- \citet[e.g.][]{thaisa2010,allan2014}) and $v_{\rm out}=80$\kms\,$/\sin\,\theta$ (value obtained from the residual map of Fig.\,\ref{res}).  \citet{pringle99} have derived the orientation of the radio jet as being $50^{\circ}$ in relative to the line of sight and therefore $\theta=40^{\circ}$. 

%to calculate the velocity of the outflow. 
%Considering the ionization cone with an opening angle of $60^{\circ}$, we find  that the front blueshifted (back redshifted) wall of the cone makes an angle of $70^{\circ}$ relative to the plane of the sky and the back blueshifted (front redshifted) wall is displaced by $10^{\circ}$ from the plane of the sky. Considering that one of the walls of the cone is almost perpendicular to the line of sight, the velocity component along the line-of-sight is expected to be very small and thus difficult to separate it from the disk component. Thus, we  interpret the redshifts to the north of the nucleus as being due to the emission of gas located at the back wall of the cone and the blueshifts to the south as being due to the front wall of the cone and therefore, we should use $\theta=70^{\circ}$ to calculate the velocity of the outflow. 
%Thus, in the residual map (Figure~\ref{res}), blueshifts and redshifts indicate the wall on the near side and the wall on the far side of the ionization cone, respectively.

The ionized gas mass outflow rate obtained under the assumptions above is $\dot{M}_{\rm{out}} \approx 0.5$~\msun\yr, being similar to the  value obtained by \citet{allan2014} using optical observations, of $\approx$0.9~\msun\yr, although these authors could not clearly resolve the geometry of the outflow due to their lower angular resolution. The obtained mass outflow rate is within the range of values obtained in previous studies of similar low-luminosity Seyfert and LINER galaxies \citep[e.g.][]{veilleux2005,crenshaw2007,barbosa2009,thaisa2010,rogemar15b}.

The mass outflow rate can also be compared to that of a compact outflow seen in hot molecular hydrogen, which is 1200 times lower, but this is because the hot molecular gas emission traces only a small fraction of the total content of molecular gas \citep{dale05,ms06,mazzalay13}. 
In any case, we note that the molecular gas outflow is oriented perpendicularly to the outflow observed in \feii\ and thus seem to have distinct origin.
As already discussed in \citet{thaisa2009}, the \feii\ is a better tracer of AGN driven outflows than the  \nii, \sii\ and even the H recombination lines.
%used by \citet{allan2014}. This may be the reason why their observations were not able to detect the biconical outflow.

%Assumimos $m_{\rm p} = $1.67\,\times\,10^{-24}\,{\rm g}$,  $N_{\rm e} = 500$\,cm$^{-3}$, $f=0.01$ e vale $v_{\rm out} = 80\,{\rm km\,s}^{-1}/sen\theta$.  
%= 2.*1.673e-24*500*80.e5*0.01*3.14*(2.25e20)**2/sin(70.*3.14/180.)*3.154e7/1.989e33 = 0.36

%= 2.*1.673e-24*500*80.e5*0.01*3.14*(2.25e20)**2/sin(40.*3.14/180.)*3.154e7/1.989e33 

Following \citet{thaisa2010}, we can use the above mass outflow rate to estimate the kinetic power of the outflow by 
\begin{equation}
\dot{E}\approx\frac{\dot{M}_{\rm out}}{2}(v^{2}_{\rm out}+\sigma^{2}), 
\end{equation} 
where $v_{\rm out}=v_{\rm obs}/\sin\,40^{\circ}$ is the velocity of the outflowing gas and $\sigma$ its velocity dispersion. Using $\sigma\approx250$\,\kms (from Figure~\ref{gas}), and $v_{\rm out} \approx125$\,\kms, we obtain $\dot{E}\approx1.3\times10^{40}$\,erg\,s$^{-1}$. The bolometric luminosity of NGC\,2110 is $L_{\rm bol}=2.9\times10^{43}$\,erg\,s$^{-1}$ \citep{allan2014}, thus the kinetic power is three orders of magnitude lower than $L_{\rm bol}$. This value is similar to those obtained for other similar targets and somewhat larger than those obtained by \citet{barbosa2009} for 6 nearby Seyfert galaxies, which are in the range $10^{-5}\leq \dot{E}/L_{\rm bol} \leq10^{-4}$. 

Therefore, only a small fraction of the mass accretion rate of $5\times10^{-3}$\,\msun\yr \citep{allan2014} 
-- necessary to power the AGN at the nucleus of NGC\,2110 -- is transformed into kinetic power in the outflow. 

%A FRASE QUE COMENTEI ESTAVA ERRADA! NO TRABALHO DA MATTEO, ELA CONCLUI QUE SOMENTE POTENCIAS MAIORES OU DA ORDEM DE 0.05 LBOL TEM FEEDBACK SIGNIFICATIVO SOBRE A GALAXIA ENTAO A CONCLUSAO TEM QUE SER OUTRA, COMO SEGUE:
This kinetic power is lower than the threshold suggested by galaxy evolution models, above which the feedback is strong enough to influence the evolution of the host galaxy \citep{dimatteo2005}.
% This value is of the order of the outflow rates due to AGN feedback derived by \citet{dimatteo2005} in simulations for co-evolution of black holes and galaxies, ensuring the $M-\sigma$ relation. 

%=0.36*1.989e33*(8500000.**2+25000000.**2)/3.154e7/2.

%=0.53*1.989e33*(12500000.**2+25000000.**2)/3.154e7/2. = 1.3055965638871E40

%The mass accretion rate necessary to power the AGN at the nucleus is  $\dot{m}=5\times10^{-3}$\msun\yr~\citep\{allan2014}. 

Although having concluded that the power of the outflow is low, our data allowed us to conclude that shocks due to the interaction of the radio jet with the galaxy circumnuclear interstellar medium  play an important role both in the kinematics and in the emission and excitation of the \feii~line.  
This is consistent with previous results from literature obtained for the central region of NGC\,2110. \citet{thaisa99}, \citet{knop2001} and \citet{ferruit2004} reported that the \feii~line exhibits a much broader and more complex profile than the \h2, being this broadening attributed to fast shocks resulting from  interaction between the radio jet and the gas in its vicinity. 

\section{Conclusions}

We have mapped the stellar population, emission-line flux distribution, gas excitation and kinematics of the inner 200~pc of the Seyfert\,2 galaxy NGC\,2110, using NIR J- and K-band  integral field spectroscopy at a spatial resolution of $\sim22$\,pc. This paper represents a continuing study of  \citet{diniz15}, where we mapped the molecular gas and stellar kinematics of the nuclear region of NGC\,2110. The main results of this paper are listed below.

\begin{itemize}
\item The young-intermediate age stellar population contribution is dominant in a ring-like structure at $\sim140$~pc from the nucleus. Similar structures have been found for other galaxies associated to low stellar velocity dispersion rings, leading to the conclusion that such rings are originated in stellar populations of ages between 100\,Myr and 1\,Gyr;

\item The distribution of the oldest stellar population dominates the inner $\sim130$~pc, inside the ring;

\item The distribution of the young stellar population (with contributions of up to 30\,\% of the continuum emission) is seen in the inner 0\farcs5 (70\,pc), being co-spatial with regions of enhanced \h2~emission, 
%***[ISTO ESTÁ ESCRITO NO PAPER, NAO VI!!! INSERIR SE NAO ESTIVER MESMO]***
indicating  that  the  molecular  gas  represents  the  circumnuclear  gas reservoir required for star formation; some recent star formation is also seen at a few locations at the external borders of the ring;
%linha 396

\item Besides the contributions of the stellar populations, the nuclear region requires additional components to fit the continuum: a FC component represented by a power-law and a hot dust component (from the dusty torus) represented by black-body components, contributing with 10\% and 35\%, respectively, to the flux at 2.2$\mu$m.
The estimated mass of hot dust in the nucleus is $\sim0.03$\msun;

\item The emission line flux distributions are distinct from each other. The \h2~emission is distributed over the whole field of view, while the H recombination lines emission is more extended along PA$\sim-30^{\circ}$. The \feii\ emission is more extended along the north--south direction, being co-spatial with the radio emission;

\item Based on emission-line ratios and the comparison of the \feii\ kinematics with the radio emission, we conclude that most of the \feii\ emission is produced by shocks due to the interaction of the radio jet with the ambient gas;

\item Distinct kinematic components are seen for distinct emission lines. The \h2~kinematics shows a rotating disk component, a compact equatorial outflow within the inner 70~pc and inflows along spiral arms. %The rotating disk component and the compact outflows ***seem also to be*** present in the H recombination lines [***TEM CERTEZA DISSO???***]. REALMENTE NÃO POSSO AFIRMAR ISSO
The \feii\ kinematics, besides presenting also the rotating disk component, shows also a biconical outflow oriented along the north--south direction, following the direction of the radio jet;

\item Further evidence of interaction of the radio jet with the \feii\ emitting gas are: an enhancement in the $\sigma$ values and line ratio values indicating shock excitation in regions surrounding the radio jet;

\item Using a kinematic model for the rotating disk component derived from the molecular gas kinematics and subtracting it from the \feii\ velocity field, we were able to isolate the outflow component. We derive an ionized gas mass outflow rate in this component of  $\dot{M}_{\rm{out}}\approx0.5$~\msun\yr~at a distance of $\sim$70~pc from the nucleus;

\item The kinetic power of the outflow is $1.3\times10^{40}$ erg\,s$^{-1}$ and is less than 0.05\% of the bolometric luminosity of the AGN in NGC\,2110, implying little influence of the corresponding feedback in the evolution of the galaxy.
\end{itemize}

\section*{Acknowledgements}

We thank the referee, Dr. Mark Durr\'e, for valuable comments which helped to improve this paper. Based on observations obtained at the Gemini Observatory, 
which is operated by the Association of Universities for Research in Astronomy,
Inc., under a cooperative agreement with the 
NSF on behalf of the Gemini partnership: the National Science Foundation (United States), the Science and Technology 
Facilities Council (United Kingdom), the National Research Council (Canada), CONICYT (Chile), the Australian Research 
Council (Australia), Minist\'erio da Ci\^encia e Tecnologia (Brazil) and South-EastCYT (Argentina).  
This research has made use of the NASA/IPAC Extragalactic Database (NED) which is operated by the Jet
 Propulsion Laboratory, California Institute of  Technology, under contract with the National Aeronautics and Space Administration.

This study was financed in part by the Coordena\c c\~ao de
Aperfei\c coamento de Pessoal de N\'ivel Superior - Brasil (CAPES) -
Finance Code 001, Conselho Nacional de Desenvolvimento Cient\'ifico e Tecnol\'ogico (CNPq) and Funda\c c\~ao de Amparo \`a pesquisa do Estado do RS (FAPERGS).

\label{lastpage}


\begin{thebibliography}{99}

%\bibitem[\protect\citeauthoryear{Afanasiev et al.}{1996}]{afanasiev96} Afanasiev, V. L., Burenkov, A. N., Shapovalova, A. I., Vlasyuk, V. V. 1996, in ASP Conf. Ser. 91, Barred Galaxies, ed. R. Buta, D. A. Crocker, B. G. Elmegreen (San Francisco, CA: ASP), 218.

%\bibitem[\protect\citeauthoryear{Antonucci}{1993}]{antonucci93} Antonucci, R.,  ARAA, 31, 473.

%\bibitem[\protect\citeauthoryear{M\"uller-S\'anchez et al.}{2011}]{sanchez2011} M\"uller-S\'anchez, F., et al. 2011, ApJ, 739, 69.

\bibitem[\protect\citeauthoryear{Asari et al.}{2007}]{asari07} Asari, N. V., Cid Fernandes, R., Stasi\'nska, G., Torres-Papaqui, J. P., Mateus, A., Sodr\'e, L., Schoenell, W. and Gomes, J. M., 2007, MNRAS, 381, 263.

%\bibitem[\protect\citeauthoryear{Barbosa et al.}{2006}]{barbosa06} Barbosa, F. K. B., Storchi-Bergmann, T., Cid Fernandes, R., Winge, C., Schmitt, H., 2006, MNRAS, 371, 170.

\bibitem[\protect\citeauthoryear{Barbosa et al.}{2009}]{barbosa2009} Barbosa, F. K. B., Storchi-Bergmann, T., Cid Fernandes, R., Winge, C. and Schmitt, H., 2009, MNRAS, 396, 2.

\bibitem[\protect\citeauthoryear{Barvainis}{1987}]{barva1987} Barvainis R., 1987, ApJ, 320, 537.

\bibitem[\protect\citeauthoryear{Blietz et al.}{1994}]{blietz1994} Blietz, M., Cameron, M., Drapatz, S., Genzel, R., Krabbe, A. and van der Werf, P. 1994, ApJ, 421, 92.

\bibitem[\protect\citeauthoryear{Bower et al.}{2006}]{bower2006} Bower R. G., Benson A. J., Malbon R., Helly J. C., Frenk C. S., Baugh C. M., Cole S. and Lacey C. G., 2006, MNRAS, 370, 645.

\bibitem[\protect\citeauthoryear{Calzetti et al.}{1994}]{calzetti94} Calzetti, D., Kinney, A.~L. and Storchi-Bergmann, T., 1994, ApJ, 429, 582.

\bibitem[\protect\citeauthoryear{Cardelli et al.}{1989}]{cardelli89} Cardelli, J. A., Clayton, G. C. and Mathis, J. S., 1989, ApJ, 345,245.

\bibitem[\protect\citeauthoryear{Cid Fernandes et al.}{2004}]{cid04} Cid Fernandes, R., Gu, Q., Melnick, J., Terlevich, E., Terlevich,  R., Kunth, D., Rodrigues Lacerda, R. and Joguet, B., 2004, MNRAS, 355, 273.

\bibitem[\protect\citeauthoryear{Cid Fernandes et al.}{2005a}]{cid05a} Cid Fernandes, R., Mateus, A., Sodr\'e L., Stasi\'nska, G. and Gomes, J. M., 2005a, MNRAS, 358, 363.

\bibitem[\protect\citeauthoryear{Cid Fernandes et al.}{2005b}]{cid05b} Cid Fernandes, R., Gonz\'alez Delgado, R. M., Storchi-Bergmann, T., Martins, L. Pires and Schmitt, H., 2005b, MNRAS, 356, 270.

\bibitem[\protect\citeauthoryear{Colina et al.}{2015}]{colina15} Colina, L. et al., 2015, A\&A, 578, 48.


\bibitem[\protect\citeauthoryear{Crenshaw et al.}{2003}]{crenshaw2003} Crenshaw, D. M., Kraemer, S. B. and Gabel, J. R., 2003, AJ, 126, 1690.

\bibitem[\protect\citeauthoryear{Crenshaw \& Kraemer}{2007}]{crenshaw2007} Crenshaw, D. M. and Kraemer, S. B., 2007, ApJ, 659, 250.

\bibitem[\protect\citeauthoryear{Dahmer-Hahn et al.}{2019}]{luis19} Dahmer-Hahn, L. G. et al., 2019, MNRAS, 482, 5211.

\bibitem[\protect\citeauthoryear{Dale et al.}{2005}]{dale05} Dale, D. A., Sheth, K., Helou, G., Regan, M. W. and H\"uttemeister, S., 2005, AJ, 129, 2197.

\bibitem[\protect\citeauthoryear{Dametto et al.}{2014}]{dametto14} Dametto, N. Z., Riffel, R., Pastoriza, M. G., Rodr\'iguez-Ardila, A., Hernandez-Jimenez, J. A. and Carvalho, E. A., 2014, MNRAS, 443, 1754.D

\bibitem[\protect\citeauthoryear{Dametto et al.}{2019}]{natacha19} Dametto, N. Z. et al. 2019, MNRAS, 482, 4437.

\bibitem[\protect\citeauthoryear{Das et al.}{2005}]{das2005} Das, V., Crenshaw, D. M., Hutchings, J. B., Deo, R. P., Kraemer, S. B., Gull, T. R., Kaiser, M. E., Nelson, C. H. and Weistrop, D., 2005, AJ, 130, 945.

\bibitem[\protect\citeauthoryear{Das et al.}{2006}]{das2006} Das, V., Crenshaw, D. M., Kraemer, S. B. and  Deo, R. P., 2006, AJ, 132, 620.


%\bibitem[\protect\citeauthoryear{Davies et al.}{2007}]{davies07} Davies, R. I., S\'anchez, F. M., Genzel, R., Tacconi, L. J., Hicks, E. K. S., Friedrich, S., Sternberg, A., 2007, ApJ, 671, 1388.

\bibitem[\protect\citeauthoryear{Davies et al.}{2009}]{davies09} Davies R. I., Maciejewski W., Hicks E. K. S., Tacconi L. J., Genzel R. and Engel H., 2009, ApJ, 702, 114D.

\bibitem[\protect\citeauthoryear{Di Matteo et al.}{2005}]{dimatteo2005} Di Matteo, T., Springel, V. and Hernquist, L., 2005, Nature, 433, 604.

\bibitem[\protect\citeauthoryear{Diniz et al.}{2015}]{diniz15}	Diniz, M. R., Riffel, R. A., Storchi-Bergmann, T. and Winge, C., 2015, MNRAS, 453, 1727.

\bibitem[\protect\citeauthoryear{Diniz et al.}{2017}]{diniz17} Diniz, M.~R., Riffel, R.~A., Riffel, R., Crenshaw, D.~M.,  Storchi-Bergmann, T., Fischer, T.~C., Schmitt, H.~R. and Kraemer, S.~B., 2017, MNRAS, 469, 3286. 

\bibitem[\protect\citeauthoryear{Diniz et al.}{2018}]{diniz18}	Diniz, M. R. et al., 2018, RNAAS, 2, 3.

\bibitem[\protect\citeauthoryear{Dors et al.}{2014}]{dors14} Dors, O. L., Cardaci, M. V., H\"agele, G. F. and Krabbe, A. C., 2014, MNRAS, 443, 1291.

\bibitem[\protect\citeauthoryear{Durr\'e \& Mould}{2014}]{durre14} Durr\'e, M. and Mould, J., 2014, ApJ, 784, 79.

\bibitem[\protect\citeauthoryear{Emsellem et al.}{2001}]{emsellem01} Emsellem, E., Greusard, D., Combes, F., Friedli, D., Leon, S., P\'econtal, E. and Wozniak, H., 2001, A\&A, 368, 52.

\bibitem[\protect\citeauthoryear{Fathi et al.}{2005}]{fathi2005} Fathi, K., van de Ven, G., Peletier, R.~F, Emsellem, E., Falc\'on-Barroso, J., Cappellari, M. and de Zeeuw, T., 2005, MNRAS, 364, 773.

%\bibitem[\protect\citeauthoryear{Deo et al.}{2006}]{deo06} Deo, R. P., Crenshaw, D. M.,  Kraemer, S. B., 2006, AJ, 132, 321.


\bibitem[\protect\citeauthoryear{Ferruit et al.}{1999}]{ferruit99}  Ferruit, P.,  Wilson, A. S., Whittle, M., Simpson, C., Mulchaey, J. S. and Ferland, G., 1999, ApJ, 523, 147. 

\bibitem[\protect\citeauthoryear{Ferruit et al.}{2004}]{ferruit2004} Ferruit P., Mundell C. G., Nagar N. M., Emsellem E., P\'econtal E., Wilson A. S. and Schinnerer, E., 2004, MNRAS, 352, 1180.

\bibitem[\protect\citeauthoryear{Forbes \& Ward}{1993}]{forbes1993} Forbes, D. A. and Ward, M. J., 1993, ApJ, 416, 150.

\bibitem[\protect\citeauthoryear{Gonz\'alez Delgado et al.}{2001}]{delgado2001} Gonz\'alez Delgado, R. M., Heckman, T. and  Leitherer, C., 2001, ApJ, 546, 845.

\bibitem[\protect\citeauthoryear{Gonz\'alez Delgado et al.}{2002}]{delgado2002} {{Gonz{\'a}lez Delgado}, R.~M., {Arribas}, S., {P{\'e}rez}, E. and {Heckman}, T.}, 2002, ApJ, 579, 188.


%\bibitem[\protect\citeauthoryear{Gonz\'alez Delgado al.}{2015}]{delgado2015} 	Gonz\'alez Delgado, R. M. et al., 2015, A\&A,  581, 44.

%\bibitem[\protect\citeauthoryear{Gonz\'alez Delgado al.}{2016}]{delgado2016} Gonz\'alez Delgado, R. M. et al., 2016, A\&A, 590, 17.

\bibitem[\protect\citeauthoryear{Hennig et al.}{2018}]{mito2018} Hennig, M.~G., Riffel, R.~A., Dors, O.~L., Riffel, R., Storchi-Bergmann, T. and Colina, L., 2018, MNRAS, 477, 1086.

\bibitem[\protect\citeauthoryear{Hicks et al.}{2009}]{hicks2009} Hicks, E. K. S., Davies, R. I., Malkan, M. A., Genzel, R., Tacconi,
L. J.; S\'anchez, F. M. and Sternberg, A., 2009, ApJ, 696, 448.

\bibitem[\protect\citeauthoryear{Hollenbach \& McKee}{1989}]{hollenbach89} Hollenbach, D. and McKee, C. F., 1989, ApJ, 342, 306.

\bibitem[\protect\citeauthoryear{Hopkins et al.}{2012}]{hopkins2012} Hopkins, P.~F., Hayward, C.~C., Narayanan, D. and Hernquist, L., 2012, MNRAS, 420, 320.

\bibitem[\protect\citeauthoryear{Imanishi \& Dudley}{2000}]{imanishi2000} Imanishi, M. and Dudley, C. C., 2000, ApJ, 545, 701.

\bibitem[\protect\citeauthoryear{Knapen et al.}{2000}]{knapen2000} Knapen, J. H., Shlosman, I. and Peletier, R. F., 2000, ApJ, 529, 93.

\bibitem[\protect\citeauthoryear{Knop et al.}{2001}]{knop2001} Knop, R. A., Armus L., Matthews K., Murphy T. W. and Soifer B. T., 2001, AJ, 122, 764.


\bibitem[\protect\citeauthoryear{Maciejewski et al.}{2002}]{mw02} Maciejewski, W. Teuben, P. J., Sparke, L. S. and Stone, J. M., 2002, MNRAS, 329, 502.

\bibitem[\protect\citeauthoryear{Maloney al.}{1996}]{maloney1996} Maloney, P. R., Hollenbach, D. J. and Tielens, A. G. G. M., 1996, ApJ, 466, 561. 


\bibitem[\protect\citeauthoryear{Maraston}{2005}]{maraston05} Maraston, C., 2005, MNRAS, 362, 799.


\bibitem[\protect\citeauthoryear{Markwardt et al.}{2009}]{mark09} Markwardt, C.~B., 2009, in Astronomical Society of the Pacific Conference Series, ed. D. A. Bohlender, D. Durand, P. Dowler, Vol. 411, 251.

%\bibitem[\protect\citeauthoryear{M\'arquez et al.}{2003}]{marquez03} M\'arquez, I., Masegosa, J., Durret, F., Gonz\'alez Delgado, R. M., Moles, M., Maza, J., P\'erez, E., Roth, M., 2003, A\&A, 409, 459.

%\bibitem[\protect\citeauthoryear{Marigo et al.}{2008}]{marigo08} Marigo P., Girardi L., Bressan A., Groenewegen M. A. T., Silva L., Granato G. L., 2008, A\&A, 482, 883.

\bibitem[\protect\citeauthoryear{Martins et al.}{2013}]{martins13} Martins, L.~P., Rodr\'iguez-Ardila, A., Diniz, S.,  Gruenwald, R. and de Souza, R., 2013, MNRAS, 431, 1823.


\bibitem[\protect\citeauthoryear{Mazzalay et al.}{2013}]{mazzalay13} Mazzalay, X. et al., 2013, MNRAS, 428, 2389.

\bibitem[\protect\citeauthoryear{McGregor et al.}{2003}]{mcgregor03} McGregor, P. J. et al., 2003, Proceedings of the SPIE, 4841, 1581.

\bibitem[\protect\citeauthoryear{Morgan}{1958}]{morgan58} Morgan W. W., 1958, PASP, 70, 364.

\bibitem[\protect\citeauthoryear{M\"uller-S\'anchez et al.}{2006}]{ms06} M\"uller-S\'anchez F., Davies R. I., Eisenhauer F., Tacconi L. J., Genzel R. and Sternberg A., 2006, A\&A, 454, 492.

\bibitem[\protect\citeauthoryear{ M\"uller-S\'anchez et al.}{2009}]{sanches2009} M\"uller-S\'anchez, F., Davies, R. I., Genzel, R., Tacconi, L. J., Eisenhauer, F., Hicks, E. K. S., Friedrich, S. and Sternberg, A., 2009, ApJ, 691, 749.


%\bibitem[\protect\citeauthoryear{Oliva et al.}{1995}]{oliva1995} Oliva, E. and Origlia, L. and Kotilainen, J.~K. and Moorwood, A.~F.~M., 1995, A\&A, 301, 55.

\bibitem[\protect\citeauthoryear{Oliva et al.}{2001}]{oliva2001} Oliva, E. et al., 2001, A\&A, 369, L5.

%\bibitem[\protect\citeauthoryear{Origlia et al.}{1993}]{origlia1993} Origlia, L. and Moorwood, A.~F.~M. and Oliva, E., 1993, A\&A, 280, 536.

\bibitem[\protect\citeauthoryear{Osterbrock \& Ferland}{2006}]{osterbrock06} Osterbrock, D. E. and Ferland, G. J., 2006, Astrophysics of Gaseous Nebulae and Active Galactic Nuclei, Second Edition, University Science Books, Mill Valley, California.

%\bibitem[\protect\citeauthoryear{Pastoriza et al.}{2009}]{pastoriza2009} Pastoriza, M.~G., Riffel, R., Rodr\'iguez-Ardila, A. and Bonatto, C., 2009, BAAA, 52, 401.

%\bibitem[\protect\citeauthoryear{P\'erez et al.}{2013}]{perez13} P\'erez E. et al., ApJ, 2013, vol. 764, 1.

\bibitem[\protect\citeauthoryear{Pringle et al.}{1999}]{pringle99} Pringle, J. E., Antonucci, R. R. J., Clarke, C. J., Kinney, A. L., Schmitt, H. R. and Ulvestad, J. S., 1999, ApJ, 526, L9.

\bibitem[\protect\citeauthoryear{Raimann et al.}{2003}]{raimann03} Raimann, D., Storchi-Bergmann, T., Gonz\'alez Delgado, R. M.,  Cid Fernandes, R., Heckman, T., Leitherer, C. and Schmitt, H. R., 2003, MNRAS, 339, 772.

\bibitem[\protect\citeauthoryear{Reunanen, Kotilainen \& Prieto}{2002}]{reunanen02} Reunanen, J., Kotilainen, J. K. and Prieto, M. A., 2002, MNRAS, 331, 154. 


\bibitem[\protect\citeauthoryear{Riffel et al.}{2006}]{rogemar2006} Riffel, R. A., Sorchi-Bergmann, T., Winge, C. and Barbosa, F. K. B., 2006, MNRAS, 373, 2.

\bibitem[\protect\citeauthoryear{Riffel et al.}{2008}]{rogemar2008} Riffel, R. A., Storchi-Bergmann, T., Winge, C., McGregor, P. J., Beck, T. and Schmitt, H., 2008, MNRAS, 385, 1129.


\bibitem[\protect\citeauthoryear{Riffel et al.}{2009a}]{rogemar2009a} Riffel, R. A., Storchi-Bergmann, T., Dors, O. L. and Winge, C., 2009a, MNRAS, 393, 783.

\bibitem[\protect\citeauthoryear{Riffel et al.}{2009b}]{rogemar2009b} Riffel, R. A., Storchi-Bergmann, T. and McGregor, P. J., 2009b, ApJ, 698, 1767.

\bibitem[\protect\citeauthoryear{Riffel}{2010a}]{rogemar2010a} Riffel, R. A., 2010a, Ap\&SS, 327, 239.

%\bibitem[\protect\citeauthoryear{Riffel et al.}{2010}]{rogemar2010b} Riffel, R. A., Storchi-Bergmann, T., Nagar, N., 2010., MNRAS, 404, 166.


\bibitem[\protect\citeauthoryear{Riffel et al.}{2010b}]{rogemar2010b} Riffel, R.~A., Storchi-Bergmann, T., Riffel, R. and Pastoriza, M.~G., 2010b, ApJ, 713, 469.



\bibitem[\protect\citeauthoryear{Riffel et al.}{2015a}]{rogemar15} Riffel, R. A. et al., 2015a, MNRAS, 446, 2823.

\bibitem[\protect\citeauthoryear{Riffel et al.}{2015b}]{rogemar15b} Riffel, R. A., Storchi-Bergmann, T. and Riffel, R., 2015b, MNRAS, 451, 3587.


\bibitem[\protect\citeauthoryear{Riffel et al.}{2016}]{rogemar16} Riffel, R. A. et al. 2016, MNRAS, 461, 4192.

\bibitem[\protect\citeauthoryear{Riffel et al.}{2017}]{rogemar2017} Riffel, R. A., Storchi-Bergmann, T., Riffel, R., Dahmer-Hahn, L. G., Diniz, M. R., Sch\"onell, A. J. and Dametto, N. Z., 2017, MNRAS, 470, 992.


\bibitem[\protect\citeauthoryear{Riffel et al.}{2018}]{rogemar2018} Riffel, R. A. et al., 2018, MNRAS, 474, 1373.

%\bibitem[\protect\citeauthoryear{Riffel et al.}{2011}]{rogemar2011a} Riffel, Rogemar A. \& Storchi-Bergmann, T., 2011, MNRAS, 411, 469.

%\bibitem[\protect\citeauthoryear{Riffel et al.}{2011}]{rogemar2011b} Riffel, R. A., Storchi-Bergmann, T., 2011b, MNRAS, 417, 2752.



%\bibitem[\protect\citeauthoryear{Riffel et al.}{2018b}]{rogemar18b} Riffel, R. A., Hekatelyne, C., Freitas, I. C., 2018, PASA, 35, 40.


%%%%%%%%%%%%%%%%%%%%

\bibitem[\protect\citeauthoryear{Riffel et al.}{2006}]{rogerio06} Riffel, R., Rodr\'iguez-Ardila, A. and Pastoriza, M. G., 2006, A\&A, 457, 61.

\bibitem[\protect\citeauthoryear{Riffel et al.}{2007}]{rogerio07} Riffel, R., Pastoriza, M. G., Rodr\'iguez-Ardila, A. and Maraston, C., 2007, ApJL, 659, 103.

\bibitem[\protect\citeauthoryear{Riffel et al.}{2008}]{rogerio2008} Riffel, R., Pastoriza, M.~G., Rodr\'iguez-Ardila, A. and Maraston, C., 2008, MNRAS, 388, 803.



\bibitem[\protect\citeauthoryear{Riffel et al.}{2009}]{rogerio2009} Riffel, R., Pastoriza, M.~G., Rodr\'iguez-Ardila, A. and Bonatto, C., 2009, MNRAS, 400, 273.

\bibitem[\protect\citeauthoryear{Riffel et al.}{2011}]{rogerio2011c} Riffel, R., Riffel, R.~A., Ferrari, F. and Storchi-Bergmann, T., 2011, MNRAS, 416, 493.

	

\bibitem[\protect\citeauthoryear{Riffel et al.}{2013}]{rogerio2013}  Riffel, R., Rodr\'iguez-Ardila, A., Aleman, I., Brotherton, M. S., Pastoriza, M. G., Bonatto, C. and Dors, O. L., 2013, MNRAS, 430, 2002.

\bibitem[\protect\citeauthoryear{Riffel et al.}{2015}]{rogerio2015} Riffel, R. et al., 2015, MNRAS, 450, 3069.

%\bibitem[\protect\citeauthoryear{Riffel et al.}{2011a}]{rogerio11a} Riffel, R., Bonatto, C. and Cid Fernandes, R. and {Pastoriza}, M.~G., Balbinot, E., 2011a, MNRAS, 411, 1897.



\bibitem[\protect\citeauthoryear{Rodr\'iguez-Ardila \& Viegas}{2003}]{ardila2003} Rodr\'iguez-Ardila, A. and Viegas, S. M., 2003, MNRAS, 340, 33.

\bibitem[\protect\citeauthoryear{Rodr\'iguez-Ardila et al.}{2004}]{ardila2004} Rodr\'iguez-Ardila, A., Pastoriza, M. G., Viegas, S. M., Sigut T. A. A. and Pradhan, A. K., 2004, A\&A, 425, 457.

\bibitem[\protect\citeauthoryear{Rodr\'iguez-Ardila et al.}{2005a}]{ardila2005} Rodr\'iguez-Ardila, A., Contini, M. and Viegas, S. M., 2005a, MNRAS, 357, 220.

\bibitem[\protect\citeauthoryear{Rodr\'iguez-Ardila et al.}{2005b}]{ardila2005b} Rodr\'iguez-Ardila, A., Riffel, R. and Pastoriza M. G., 2005b, MNRAS, 364, 1041.


\bibitem[\protect\citeauthoryear{Rodr\'igues-Ardila \& Mazzalay}{2006}]{ardila2006} Rodr\'iguez-Ardila, A. and Mazzalay, X., 2006, MNRAS, 367, 57.


\bibitem[\protect\citeauthoryear{Rosario et al.}{2010}]{rosario2010} Rosario D. J., Whittle M., Nelson C. H. and Wilson A. S., 2010, MNRAS, 408, 565.

\bibitem[\protect\citeauthoryear{Rosario et al.}{2019}]{rosario19} Rosario, D. J., Togi, A., Burtscher, L., Davies, R., Shimizu, T., and Lutz, D., 2019, arXiv:1903.07637

\bibitem[\protect\citeauthoryear{Schartmann et al.}{2010}]{schartmann2010} Schartmann, M., Burkert, A., Krause, M., Camenzind, M., Meisenheimer, K. and Davies, R.~I., 2010, MNRAS, 403, 1801.


\bibitem[\protect\citeauthoryear{Schartmann et al.}{2014}]{schartmann2014} Schartmann, M., Wada, K., Prieto, M.~A., Burkert, A. and Tristram, K.~R.~W., 2014, MNRAS, 445, 3878.

\bibitem[\protect\citeauthoryear{Sersic \& Pastoriza}{1967}]{sersic67} Sersic, J. L.  and Pastoriza, M., 1967, PASP, 79, 152.

\bibitem[\protect\citeauthoryear{Schnorr-M\"uller et al.}{2014}]{allan2014} Schnorr-M\"uller, A., Storchi-Bergmann, T.,  Nagar, N.~M., Robinson, A. Lena, D., Riffel, R.~A. and Couto, G.~S., 2014, MNRAS, 437, 1708.

\bibitem[\protect\citeauthoryear{Sch\"onell et al.}{2017}]{astor17} Sch\"onell, Jr., A.~J., Storchi-Bergmann, T., Riffel, R.~A. and Riffel, R., 2017, MNRAS, 464, 1771.


\bibitem[\protect\citeauthoryear{Simpson et al.}{1996}]{simpson96}  Simpson, C., Forbes, D. A., Baker, A. C. and Ward, M. J., 1996, MNRAS, 283, 777.


\bibitem[\protect\citeauthoryear{Somerville et al.}{2008}]{somerville2008} Somerville R. S., Hopkins P. F., Cox T. J., Robertson B. E. and Hernquist L., 2008, MNRAS, 391, 481.



%\bibitem[\protect\citeauthoryear{Calzetti et al.}{2000}]{calzetti2000} Calzetti, D. and Armus, L. and Bohlin, R.~C. and Kinney, A.~L. and Koornneef, J. and Storchi-Bergmann, T., 2000, ApJ, 533, 682.

\bibitem[\protect\citeauthoryear{Storchi-Bergmann et al.}{1999}]{thaisa99} Storchi-Bergmann,  T.,  Winge,  C.,  Ward,  M.  J. and   Wilson,  A.  S.,  1999,  MNRAS, 304, 35.

\bibitem[\protect\citeauthoryear{Storchi-Bergmann et al.}{2009}]{thaisa2009} Storchi-Bergmann T., McGregor P., Riffel R. A., Sim\~oes Lopes R., Beck T. and Dopita M., 2009, MNRAS, 394, 1148.

\bibitem[\protect\citeauthoryear{Storchi-Bergmann et al.}{2010}]{thaisa2010} Storchi-Bergmann, T., Sim\~oes Lopes, R.,McGregor, P. Riffel, R. A., Beck, T. and Martini, P., 2010, MNRAS, 402, 819.

\bibitem[\protect\citeauthoryear{Storchi-Bergmann et al.}{2012}]{ngc1068} Storchi-Bergmann, T., Riffel, R. A., Riffel, R., Diniz, M. R., Borges Vale, T. and McGregor, P. J., 2012, ApJ, 755, 87.

\bibitem[\protect\citeauthoryear{Storchi-Bergmann \& Schnorr-M\"uller}{2019}]{thaisa19} Storchi-Bergmann, T. and  Schnorr-M\"uller, A., 2019, NatAs, 3, 48. 

\bibitem[\protect\citeauthoryear{Ulvestad \& Wilson}{1984}]{ulvestad84} Ulvestad, J. S. and Wilson, A. S., 1984, ApJ, 285, 439.

\bibitem[\protect\citeauthoryear{Veilleux et al.}{2005}]{veilleux2005} Veilleux S., Cecil G. and Bland-Hawthorn J., 2005, ARA\&A, 43, 769.

\bibitem[\protect\citeauthoryear{Zibetti et al.}{2013}]{zibetti2013} Zibetti, S.,  Gallazzi, A., Charlot, S., Pierini, D. and Pasquali, A., 2013, MNRAS, 428, 1479.

%%%%%%%%%%%%%%%%%%%%%%%%%%%%%%%%%%%%%%%%%%%%%%%%%%%%%%%%%




%\bibitem[\protect\citeauthoryear{Dors et al.}{2008}]{dors08} Dors, O. L., Storchi-Bergmann, T., Riffel, Rogemar A., Schmidt, A., 2008, A\&A, 482, 59.


%\bibitem[\protect\citeauthoryear{Fathi et al.}{2006}]{fathi2006} Fathi, K., Storchi-Bergmann, T., Riffel, R. A., Winge, C., Axon, D. J., Robinson, A., Capetti, A., Marconi, A., 2006, ApJ, 641, 25.


%\bibitem[\protect\citeauthoryear{Ferruit et al.}{2000}]{ferruit2000} Ferruit, P., Wilson, A. S., Mulchaey, J. 2000, ApJS, 128, 139.

%\bibitem[\protect\citeauthoryear{Fischer et al.}{2017}]{fischer17} Fischer T. C., C. Machuca, M. R. Diniz, D. M. Crenshaw, S. B. Kraemer, R. A. Riffel, H. R. Schmitt, F. Baron, T. Storchi-Bergmann, A. Straughn, M. Revalski, C. Pope, ApJ, 834, 30.

%\bibitem[\protect\citeauthoryear{Fischer et al.}{2015}]{fischer15} Fischer, T. C., Crenshaw, D. M., Kraemer, S. B., Schmitt, H. R., Storchi-Bergmann, T., Riffel, R. A., 2015, ApJ, 799, 234.



%\bibitem[\protect\citeauthoryear{Fischer et al.}{2010}]{fischer2010} Fischer T. C., D. M. Crenshaw, S. B. Kraemer, H. R. Schmitt, M. L. Trippe, 2010, AJ, 140, 577.https://www.overleaf.com/18492155vfbkgfjzqdpk

%\bibitem[\protect\citeauthoryear{Garcia-Vargas et al.}{1989}]{garcia89} Garcia-Vargas, M. L.; Diaz, A. I.; Terlevich, R.; Terlevich, E., 1989, Ap\&SS, 157, 125.




%\bibitem[\protect\citeauthoryear{Haniff et al.}{1988}]{haniff88} Haniff, C.A., Wilson, A.S., Ward, M.J., 1988, ApJ 334, 104.

%\bibitem[\protect\citeauthoryear{Heckman et al.}{1997}]{heckman97} Heckman, T. M., Gonz\'alez Delgado, R. M., Leitherer, C., Meurer, G. R., Krolik, J., Wilson, A. S., Koratkar, A., Kinney, A. 1997, ApJ, 482, 114.



%\bibitem[\protect\citeauthoryear{Imanishi}{2002}]{imanishi2002} Imanishi, M., 2002, ApJ, 569, 44.



%\bibitem[\protect\citeauthoryear{Maciejewski}{2004a}]{mw04a} Maciejewski, W., 2004a, MNRAS, 654, 883.

%\bibitem[\protect\citeauthoryear{Maciejewski}{2004b}]{mw04b} Maciejewski, W., 2004b, MNRAS, 654, 892.


%\bibitem[\protect\citeauthoryear{Maraston C. et al.}{2006}]{maraston16} Maraston C., Daddi E., Renzini A., Cimatti A., Dickinson M., Papovich C., Pasquali A., Pirzkal N., 2006, ApJ, 652, 85.








%\bibitem[\protect\citeauthoryear{Norman \& Scoville}{1988}]{norman88} Norman, C., \& Scoville, N. 1988, ApJ, 332, 124.




%\bibitem[\protect\citeauthoryear{Perry \& Dyson}{1985}]{perry-dyson85} Perry, J. J., \& Dyson, J. E. 1985 MNRAS, 213, 665.

%\bibitem[\protect\citeauthoryear{Plummer}{1911}]{plum} Plummer, H.~C., 1911, ApJ, 71, 460. 



%\bibitem[\protect\citeauthoryear{Ramos Almeida et al.}{2008}]{ramos2008} Ramos Almeida C., A. M. P\'erez Garc\'ia, J. A. Acosta-Pulido, and O. G\'onzalez-Mart\'in, 2008, ApJL, 680, 17. 

%\bibitem[\protect\citeauthoryear{Ramos Almeida,  P\'erez Garc\'ia \& Acosta-Pulido}{2009}]{ramos2009}  Ramos Almeida, C., P\'erez Garc\'ia,  A. M. , Acosta-Pulido,  J. A., 2009, ApJ, 694, 1379. 

%\bibitem[\protect\citeauthoryear{Revalski et al.}{2018}]{revalski2018}  Revalski M., Crenshaw D. M., Kraemer S. B., Fischer T. C., Schmitt H. R., Machuca C., 2018, ApJ, 856, 46.




%\bibitem[\protect\citeauthoryear{Riffel, Storchi-Bergmann \& Winge}{2013}]{rogemar2013} Riffel, R. A., Storchi-Bergmann, T., Winge, C., 2013, 430, 2249.

%\bibitem[\protect\citeauthoryear{Salaris et al.}{2014}]{salaris14} Salaris, M., Weiss, A., Cassar\`a, L. P., Piovan, L., Chiosi, C., 2014, A\&A, 565.

%\bibitem[\protect\citeauthoryear{S\'anchez et al.}{2016}]{sanchez2016}	S\'anchez, S. F.; P\'erez, E.; S\'anchez-Bl\'azquez, P.; Garc\'ia-Benito, R.; Ibarra-Mede, H. J.; Gonz\'alez, J. J.; Rosales-Ortega, F. F.; S\'anchez-Menguiano, L.; Ascasibar, Y.; Bitsakis, T.; Law, D.; Cano-D\'iaz, M.; L\'opez-Cob\'a, C.; Marino, R. A.; Gil de Paz, A.; L\'opez-S\'anchez, A. R.; Barrera-Ballesteros, J.; Galbany, L.; Mast, D.; Abril-Melgarejo, V.; Roman-Lopes, A., 2016, RMXAA, 52, 171.

%\bibitem[\protect\citeauthoryear{P\'erez et al.}{2013}]{perez13} P\'erez E., Cid Fernandes R., Gonz\'alez Delgado R. M., Garc\'ia-Benito R., S\'anchez S. F.,Husemann B., Mast D., Rod\'on J. R., Kupko D., Backsmann N., de Amorim A. L., van de Ven G., Walcher J., ApJ, 2013, vol. 764, 1.

%\bibitem[\protect\citeauthoryear{Delgado al.}{2015}]{delgado2015} 	Gonz\'alez Delgado, R. M.; Garc\'ia-Benito, R.; P\'erez, E.; Cid Fernandes, R.; de Amorim, A. L.; Cortijo-Ferrero, C.; Lacerda, E. A. D.; L\'opez Fern\'andez, R.; Vale-Asari, N.; S\'anchez, S. F.; Moll\'a, M.; Ruiz-Lara, T.; S\'anchez-Bl\'azquez, P.; Walcher, C. J.; Alves, J.; Aguerri, J. A. L.; Bekerait\'e, S.; Bland-Hawthorn, J.; Galbany, L.; Gallazzi, A.; Husemann, B.; Iglesias-P\'aramo, J.; Kalinova, V.; L\'opez-S\'anchez, A. R.; Marino, R. A.; M\'arquez, I.; Masegosa, J.; Mast, D.; M\'endez-Abreu, J.; Mendoza, A.; del Olmo, A.; P\'erez, I.; Quirrenbach, A.; Zibetti, S., 2015, A\&A,  581, 44.

%\bibitem[\protect\citeauthoryear{Delgado al.}{2016}]{delgado2016} Gonz\'alez Delgado, R. M.; Cid Fernandes, R.; P\'erez, E.; Garc\'ia-Benito, R.; L\'opez Fern\'andez, R.; Lacerda, E. A. D.; Cortijo-Ferrero, C.; de Amorim, A. L.; Vale Asari, N.; S\'anchez, S. F.; Walcher, C. J.; Wisotzki, L.; Mast, D.; Alves, J.; Ascasibar, Y.; Bland-Hawthorn, J.; Galbany, L.; Kennicutt, R. C.; M\'arquez, I.; Masegosa, J.; Moll\'a, M.; S\'anchez-Bl\'azquez, P.; V\'ilchez, J. M., 2016, A\&A, 590, 17.

%\bibitem[\protect\citeauthoryear{Sarzi et al.}{2007}]{sarzi07} Sarzi, M., Allard, E.~L., Knapen, J.~H., Mazzuca, L.~M., 2007, MNRAS, 380, 949.

%\bibitem[\protect\citeauthoryear{Schmitt et al.}{1999}]{schmitt99} Schmitt H.R., Storchi-Bergmann T., Cid Fernandes R., 1999, MNRAS, 303, 173.


%\bibitem[\protect\citeauthoryear{Schlesinger et al.}{2009}]{schlesinger09} Schlesinger, K., Pogge, R.~W., Martini, P., Shields, J.~C., Fields, D., 2009, ApJ, 699, 857.




%\bibitem[\protect\citeauthoryear{Sch\"onell et al.}{2016}]{astor16} Sch\"onell, A. J., Storchi-Bergmann, T., Riffel, R. A., Riffel, R., 2016, MNRAS, tmp1370.

%\bibitem[\protect\citeauthoryear{Sch{\"o}nell et al.}{2014}]{astor2014} Sch\"onell, A.~J., Riffel, R.~A., Storchi-Bergmann, T., Winge, C., 2014, MNRAS, 445, 414.

%\bibitem[\protect\citeauthoryear{Springob et al.}{2005}]{springob05} Springo, C. M., Haynes, M. P., Giovanelli, R., Kent, B. R. 2005, ApJS, 160, 149.

%\bibitem[\protect\citeauthoryear{Sim\~oes Lopes et al.}{2006}]{simoes07} Sim\~oes Lopes, R. D., Storchi-Bergmann, T., de F´atima Saraiva, M., Martini, P., 2007, ApJ, 655, 718.


%\bibitem[\protect\citeauthoryear{Storchi-Bergmann et al.}{1996}]{thaisa96} Storchi-Bergmann, T., Wilson, A. S., Mulchaey, J. S., Binette, L. 1996, A\&A, 312, 357.

%\bibitem[\protect\citeauthoryear{Storchi-Bergmann et al.}{2001}]{sb01} Storchi-Bergmann, T., Gonz\'alez Delgado, R. M., Schmitt, H. R., Cid Fernandes, R., Heckman, T., 2001, ApJ, 559, 147.

%\bibitem[\protect\citeauthoryear{Terlevich \& Melnick}{1985}]{terlevich-melnick85} Terlevich, R., \& Melnick, J. 1985, MNRAS, 213, 841.

%\bibitem[\protect\citeauthoryear{Terlevich et al.}{1990}]{terlevich90} Terlevich, E., Diaz, A. I.,  Terlevich, R. 1990, MNRAS, 242, 271.

%\bibitem[\protect\citeauthoryear{Tsvetanov et al.}{1992}]{tsvetanov92} Tsvetanov, Z., Walsh, J. R. 1992, ApJ, 386, 485.

%\bibitem[\protect\citeauthoryear{Unger et al.}{1987}]{unger87} Unger, S. W., Pedlar, A., Axon, D. J., Whittle, M., Meurs, E. J. A., Ward, M. J., 1987, MNRAS, 228, 671.

%\bibitem[\protect\citeauthoryear{Urry \& Padovani}{1995}]{urry95} Urry, C. M., Padovani, P., 1995, PASP, 107, 803.

%\bibitem[\protect\citeauthoryear{Veilleux et al.}{1997}]{veilleux97} Veilleux, S., Goodrich, R. W., Hill, G. J., 1997, ApJ, 477, 631.


%\bibitem[\protect\citeauthoryear{de Vaucouleurs et al.}{1991}]{vaucoleurs91} de Vaucouleurs, G., de Vaucouleurs, A., Corwin, Jr., H.~G., Buta, R.~J., Paturel, G., Fouqu\'e, P. 1991, Third Reference Catalogue of Bright Galaxies RC3) Springer Verlag, New York.

%\bibitem[\protect\citeauthoryear{Winge et al.}{2009}]{winge09} Winge, C., Riffel, R., A., Storchi-Bergmann T., 2009, ApJS, 185, 186.





%\bibitem[\protect\citeauthoryear{Bica \&  Alloin}{1986}]{bica1986} Bica, E. and Alloin, D., 1986, A\&A, 162, 21.

%\bibitem[\protect\citeauthoryear{Schmitt et al.}{1996}]{schmitt1996} Schmitt, H.~R. and Bica, E. and Pastoriza, M.~G., 1996, MNRAS, 278, 965.

%\bibitem[\protect\citeauthoryear{Schmitt et al.}{1996}]{schmitt1996b} Schmitt H. R., Kinney A. L., 1996, ApJ, 463, 498. 


%\bibitem[\protect\citeauthoryear{Schmitt et al.}{2000}]{raimann2000a} Raimann, D. and Bica, E. and Storchi-Bergmann, T. and Melnick, J. and Schmitt, H., 2000, MNRAS, 314, 295.

%\bibitem[\protect\citeauthoryear{Schmitt et al.}{2000}]{raimann2000b} Raimann, D. and Storchi-Bergmann, T. and Bica, E. and Melnick, J. and Schmitt, H., 2000, MNRAS, 316, 559.

%\bibitem[\protect\citeauthoryear{Rembold et al.}{2007}]{rembold2007} Rembold, S.~B. and Pastoriza, M.~G., 2007, MNRAS, 374, 1056.



%\bibitem[\protect\citeauthoryear{Melbourne et al.}{2012}]{melbourne12} Melbourne J. and Williams, B.~F. and Dalcanton, J.~J. and Rosenfield, P. and Girardi, L. and Marigo, P. and Weisz, D. and Dolphin, A. and Boyer, M.~L. and Olsen, K. and Skillman, E. and Seth, A.~C., 2012, ApJ, 748, 47.


%\bibitem[\protect\citeauthoryear{Riffel et al.}{2011}]{rogerio2011a} Riffel, R. and Ruschel-Dutra, D. and Pastoriza, M.~G. and Rodr\'iguez-Ardila, A. and Santos, Jr., J.~F.~C. and Bonatto, C.~J. and Ducati, J.~R., 2011, MNRAS, 410, 2714.


%\bibitem[\protect\citeauthoryear{V\'eron-Cetty \& V\'eron, P.}{2006}]{veron2006} V\'eron-Cetty, M.-P. \& V\'eron, P. 2006,  A\&A, 455, 773.

%\bibitem[\protect\citeauthoryear{Wilson et al.}{1993}]{wilson1993} Wilson A. S., Braatz J. A., Heckman T. M., Krolic J. H., Miley G. K., 1993, ApJ, 419, L61.




%\bibitem[\protect\citeauthoryear{Bertola et al.}{1991}]{bertola91} Bertola, F., Bettoni, D., Danziger, J., Sadler, E., Sparke \& L., de Zeeuw, T., 1991, ApJ, 373, 369.


%\bibitem[\protect\citeauthoryear{Boisson et al.}{2000}]{boisson2000}  {{Boisson}, C. and {Joly}, M. and {Moultaka}, J. and {Pelat}, D. and 	{Serote Roos}, M.}, 2000, MNRAS, 357, 850.




%\bibitem[\protect\citeauthoryear{Bruzual et al.}{2003}]{bruzual03} Bruzual G., Charlot S., 2003, MNRAS, 344, 1000.

%\bibitem[\protect\citeauthoryear{Calzetti et al.}{1994}]{calzetti94}  Calzetti, D., Kinney, A. L., Storchi-Bergmann, T. 1994, ApJ, 429, 582.

%\bibitem[\protect\citeauthoryear{Cappellari \& Emsellem}{2004}]{cappellari04} Cappellari, M., Emsellem, E., 2004, PASP, 116, 138.

%\bibitem[\protect\citeauthoryear{Capozzi et al.}{2016}]{capozzi16} Capozzi D., Maraston C., Daddi E., Renzini A., Strazzullo V. , Gobat R., 2016, MNRAS 456, 790.

\end{thebibliography}
\end{document}